\documentclass[twocolumn,amsmath,amssymb,prl,longbibliography]{revtex4-1}
\usepackage{multirow}
\usepackage{array}
\usepackage{amsmath}
\usepackage{graphicx}
\usepackage{dcolumn}
\usepackage{bm}
\DeclareMathAlphabet \mathbfcal{OMS}{cmsy}{b}{n}
\begin{document}



\title{Femtosecond valley polarization and topological resonances in transition metal dichalcogenides}

\author{S. Azar Oliaei Motlagh}
 \author{Jhih-Sheng Wu}
\author{Vadym Apalkov}
\author{Mark I. Stockman}
\affiliation{Center for Nano-Optics (CeNO) and
Department of Physics and Astronomy, Georgia State
University, Atlanta, Georgia 30303, USA
}

\date{\today}
\begin{abstract}
We theoretically introduce the fundamentally fastest induction of a significant population and valley polarization in a monolayer of a transition metal dichalcogenide (i.e., $\mathrm{MoS_2}$ and $\mathrm{WS_2}$). This may be extended to other two-dimensional materials with the same symmetry. This valley polarization can be written and read-out by a pulse consisting of just a single optical oscillation with  a duration of a few femtoseconds and an  amplitude of $\sim0.2~\mathrm V/\mathrm{\AA}$.  Under these conditions, we predict a new effect of {\em topological resonance}, which is due to  Bloch motion of electrons in the reciprocal space where electron population textures are formed defined by  non-Abelian Berry curvature. The predicted phenomena can be applied for information storage and processing in PHz-band optoelectronics.
\end{abstract}
\maketitle

Femtosecond and attosecond technology has made it possible to control and study ultrafast electron dynamics in three-dimensional solids  \cite{Fattahi_Third-generation_femtosecond_technology_Optica_2014, Schiffrin_at_al_Nature_2012_Current_in_Dielectric, Schultze_et_al_Nature_2012_Controlling_Dielectrics, Baltuska_et_al_Attosecond_IonizationPRL_2011, Goulielmakis_et_al_Lightwave_Electronics_Science_2007, Paasch_Colberg_et_al_Optica_2016_optical_control}. There is also a wide class of two-dimensional (2D) crystals that, in particular, can be obtained by exfoliation from layered materials, which have unique and useful properties \cite{Strano_et_al_nnano.2012.193_Transitional_Metal, Novoselov_et_al_Science_2013_Light_Interaction_with_2D_Materials, Novoselov_et_al_Nat_Mat_2015_LED_of_2D_Heterostructures, Liu_et_al_Chemical_Society_Rev_2015_Electronic_structures, Novoselov_et_al_Science_2016_2D_materials_and_van_der_Waals,  Ye_et_al_Nature_Nanotechnology_2016_Electrical_generation_and_control, Reis_et_al_Nat_Phys_2017_HHG_from_2D_Crystals, Ashton_et_al_PRL_2017_Topology_Scaling_Identification}. 
This is a modern class of materials bearing a promise for applications in ultrafast opto-electronics  \cite{lemme_li_palacios_schwierz_2014}. However, not all 2D materials are suitable for any given application. 
For example, graphene is a well-studied 2D material with many interesting and useful properties. However, it is semimetallic with no bandgap between the valence band (VB) and the conduction band (CB). Consequently, a graphene transitor exhibits a relatively high off-current, which drastically limits its usefulness 
 \cite{Novoselov_et_al_Nature_2012_Graphene_Review, Jariwala_et_al_Asc_Nano_2014_Transition_Metal, Geim_et_al_Nat_Mater_2007_The_rise_of_graphene, Schwierz_Nature_Nanotechnology_2010_Graphene_transistors}. In contrast to graphene, there is a broad class of 2D semiconductors possessing finite bandgaps. Among them, transition metal dichalcogenides (TMDCs) possess bandgaps of $1.1-2.1$ eV  \cite{Strano_et_al_nnano.2012.193_Transitional_Metal, Liu_et_al_PRB_2014_Three_Band_Model, Liu_et_al_Chemical_Society_Rev_2015_Electronic_structures, Novoselov_et_al_Science_2016_2D_materials_and_van_der_Waals, Xiao_et_al_PRL_2012_Coupled_Spin_and_Valley_Physics, Jiang_Frontiers_of_Physics_2015_Graphene_versus_MoS2}.

Similar to graphene, TMDC monolayers have hexagonal lattices constituted by two triangular sublattices 
 \cite{Geim_et_al_Nat_Mater_2007_The_rise_of_graphene, Liu_et_al_PRB_2014_Three_Band_Model, Liu_et_al_Chemical_Society_Rev_2015_Electronic_structures}. However, unlike graphene, these sublattices consist of different atoms (metal and chalcogen), which breaks the inversion ($\cal P$) symmetry and opens up gaps at the $K, K^\prime$-points. The degeneracy of the the  $K$- and $K^\prime$-valleys \ is protected by the time reversal ($\cal T$) symmetry \cite{Geim_et_al_Nat_Mater_2007_The_rise_of_graphene, Sie_et_al_Nature_Materials_2015_Valley-selective_optical}.
 
The $\cal T$-symmetry and valley degeneracy can be relaxed by circularly-polarized optical pumping, which allows for a highly valley-specific electron population, depending on the helicity of the excitation pulse   \cite{Xiao_et_al_PRL_2012_Coupled_Spin_and_Valley_Physics, Zeng_et_al_Nature_Nanotechnology_2012_Valley_polarization_in_MoS2, Heinz_et_al_10.1038_Nnano.2012.96_Valley_Polarization_in_TMDC_by_Optical_Helicity, Feng_et_al_ncomms1882_2012_Valley_Selective_CD, Jones_et_al_Nature_Nanotechnology_2013_Optical_generation_of_excitonic_valley, Sie_et_al_Nature_Materials_2015_Valley-selective_optical}.  This selective valley population, known as valley polarization, introduces a new area referred to as valleytronics  \cite{Eginligil_et_al_Nature_Communications_2015_Dichroic_spin_valley_photocurrent, Ye_et_al_Nature_Nanotechnology_2016_Electrical_generation_and_control}. In addition to the valley degree of freedom of  TMDC monolayers, a significant spin-orbit coupling (SOC) makes these materials promising also for spintronics  \cite{Xiao_et_al_PRL_2012_Coupled_Spin_and_Valley_Physics, Heinz_et_al_10.1038_Nnano.2012.96_Valley_Polarization_in_TMDC_by_Optical_Helicity}.  

\begin{figure}
\begin{center}\includegraphics[width=0.47\textwidth]{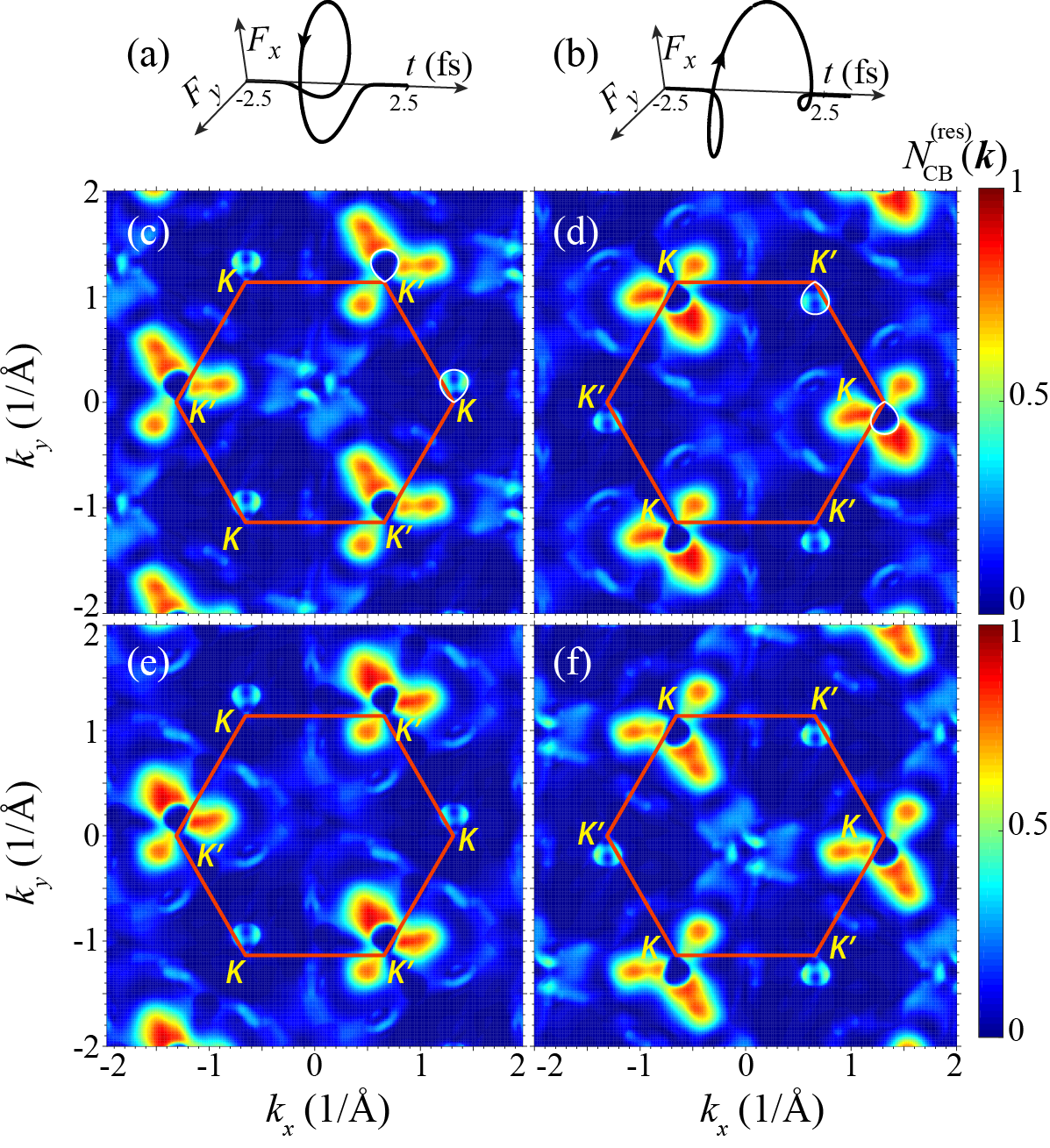}\end{center}
\caption{(Color online) Residual CB population $N\mathrm{^{(res)}_\mathrm{CB}}(\mathbf{k})$ for monolayer $\mathrm{MoS_2}$ in the extended zone picture. The red solid line shows the first Brillouin zone boundary with $K, K^\prime$-points indicated. The amplitude of the optical field is 0.2 $\mathrm{V\AA^{-1}}$. (a)  Waveform $\mathbf F(t)$ for right-handed circularly-polarized pulse.  (b) The same as panel (a) but for left-handed circularly-polarized pulse. (c) Residual population of spin-up electrons, $N\mathrm{^{(res)}_\mathrm{CB\uparrow}}(\mathbf{k})$, for right-handed pulse. (d) The same as (c), $N\mathrm{^{(res)}_\mathrm{CB\uparrow}}(\mathbf{k})$, but for left-handed pulse. (e)  Residual population of spin-down electrons, $N\mathrm{^{(res)}_\mathrm{CB\downarrow}}(\mathbf{k})$, for right-handed pulse. (f) The same as (e), $N\mathrm{^{(res)}_\mathrm{CB\downarrow}}(\mathbf{k})$, but for left-handed pulse.
}  
\label{MoS2}
\end{figure}

In this Letter, we theoretically introduce the fundamentally fastest induction of a significant valley polarization in $\mathrm{MoS_2}$ and $\mathrm{WS_2}$ monolayers by a {\em single} cycle of a strong circularly-polarized optical field with duration of a few fs and amplitude of $F_0=0.2-0.5~\mathrm{V\AA^{-1}}$. This process is determined by a strong-field-induced electron motion in the reciprocal space, spanning a significant part of the Brillouin zone. This motion also causes a new effect,  {\em topological resonance}, which we introduce below in discussion of Fig.\ \ref{MoS2}.

For a single-oscillation pulse, optical electric field $\mathbf F(t)$ depending on time $t$ is defined as
\begin{equation}
F_x(t)= F_0(1-2u^2)e^{-u^2}~,~~~F_y(t)=\pm2uF_0e^{-u^2}~,
\label{Field}
\end{equation}
where $u=t/\tau$, and $\tau=1$ fs determines the pulse duration and its mean frequency (see appendix for definition), $\hbar\bar\omega\approx1.2$ eV. The $\pm$ sign defines helicity of the applied pulse: $+$ for the right-handed and $-$ for the left-handed circular polarization. Defined by Eq.\ (\ref{Field}), these left- and right-handed pulses are $\cal T$-reversed with respect to one another. A few- or single-oscillation  pulses are presently readily available from near-ultraviolet through terahertz range in linear \cite{Garg-Nature_2016-Multi-petahertz_electronic_metrology, Krogen-2017-Generation_and_multi-octave_shaping_in_mid-IR, Liang-Nat_Commun_2017-High-energy_mid-infrared_sub-cycle, Russel_et_al_optica-4-9-1024_2017_Single_Cycle_MidIR_Pulses_from_OPCPA, Hauri_PhysRevLett.118_2017_Subcycle_THz_Semiconductors, Dhillon_2017_terahertz_science_roadmap, Langer-2017-Symmetry-controlled_THz} or circularly polarization  \cite{Dhillon_2017_terahertz_science_roadmap, Li-2016-Polarization_gating_of_HHG}.

 We set the TMDC monolayer in the $xy$ plane with the pulse incident in the $z$ direction. We use a three-band tight binding (TB) (third nearest neighbor) model Hamiltonian \cite{Liu_et_al_PRB_2014_Three_Band_Model}, $H^\mathrm{TNN}$, see Eq. (8) of appendix. Unlike the TB model of graphene, which is constructed of a single orbital per sublattice, the TB Hamiltonian of a TMDC monolayer  is constituted by three orbitals, $d_{z^2}$, $d_{xy}$, and $d_{x^2-y^2}$ of the metal atom. The full Hamiltonian is $H=H^\mathrm{TNN}+H^\mathrm{SOC}+H^\mathrm{int}(t)$, where $H^\mathrm{SOC}$ is the SOC term [Eq.\ (10) of appendix], and $H^\mathrm{int}(t)$ is the light-TMDC interaction term. The latter we write down  in the length gauge: $H^\mathrm{int}=e\mathbf F(t)\mathbf r$, where $e$ is unit charge. 

We assume that during the excitation pulse, electron dynamics is Hamiltonian (coherent), and electron collisions can be neglected. This  is a valid assumption since the applied pulse (a few femtoseconds) is much shorter than the electron scattering (dephasing) time in TMDCs. In fact, this dephasing time was reported to be 500 fs for an  atomicaly thin $\mathrm{MoS_2}$  \cite{Sim_et_al_PRB_2013_Exciton_dynamics_in_atomically_thin_MoS2}. Also, Ref.\ \onlinecite{Moody_et_al_Journal_of_the_Optical_Society_2016_Exciton_dynamics_in_monolayer} reported electron coherence times for $\mathrm{WSe_2}$ to be in the interval of $150-410$ fs. Additionally, in Ref.\ \onlinecite{Nie_et_al_Acs_Nano_2014_Ultrafast_Carrier_Thermalization}, the time of dephasing was calculated theoretically to be $\approx37$ fs for a few layers of  $\mathrm{MoS_2}$. Free carrier relaxation time in $\mathrm{MoS_2}$ was found to be 25 ps, and the electron-hole recombination time to be 300 ps \cite{Wang-2015-Ultrafast_Multi-Level_Logic_Gates}.
Based on these arguments, we describe the electron dynamics as coherent by time-dependent Schr\"odinger equation (TDSE). Previously, such a TDSE theory \cite{Apalkov_Stockman_PRB_2012_Strong_Field_Reflection, Apalkov_Stockman_PRB_2013_Metal_Nanofilm_in_Ultratsrong_Fields, Stockman_et_al_PhysRevLett.113_2014_HHG, Stockman_et_al_PhysRevB.90_2014_WS_States_of_Graphene, Stockman_et_al_PRB_2016_Graphene_in_Ultrafast_Field, Stockman_PhysRevB.92_2015_Ultrafast_Control_Symmetry, Stockman_et_al_PhysRevB.93.155434_Graphene_Circular_Interferometry, Stockman_Yakovlev_et_al_PhysRevLett.116_2016_Strong_Field_Dynamics_in_Semiconductors, Stockman_et_al_PhysRevB.95_2017_Crystalline_TI} was successful in predicting new effects, stimulating experimental research, and describing expermental results in both three-dimensional solids  \cite{Schiffrin_at_al_Nature_2012_Current_in_Dielectric, Schultze_et_al_Nature_2012_Controlling_Dielectrics, Stockman_et_al_Sci_Rep_2016_Semimetallization} and graphene \cite{Higuchi_Hommelhoff_et_al_Nature_2017_Currents_in_Graphene}. For non-interacting particles, the TDSE theory is fundamentally equivalent to the corresponding density matrix equations
but is  computationally much more efficient.

We will employ an interaction representation in an adiabatic basis of the Houston functions \cite{Houston_PR_1940_Electron_Acceleration_in_Lattice}, $\Phi^\mathrm{(H)}_{\alpha {\bf q}}({\bf r},t)$, which exactly takes into account the intraband (adiabatic) electron dynamics. Then, the Hamiltonian has only off-diagonal matrix elements describing interband transitions. 
A general solution of TDSE is 
\begin{equation}
\Psi_{\bf q} ({\bf r},t)=\sum_{\alpha=c_1,c_2,v}\beta_{\alpha{\bf q}}(t) \Phi^\mathrm{(H)}_{\alpha {\bf q}}({\bf r},t),
\end{equation}
where $v,c_1,c_2$ denote the highest valence band and the two lowest conduction bands, respectively;
$\beta_{\alpha{\bf q}}(t)$ are expansion coefficients satisfying equations
\begin{equation}
\frac{{d{\beta _{\alpha{\bf{q}}}}(t)}}{{dt}} =  - \frac{i}{\hbar }\sum_{\alpha_1\neq \alpha}\mathbf F(t) \mathbf Q_{\alpha\alpha_1}(\mathbf q, t) {\beta _{\alpha_1 {\bf{q}}}}(t) ,
\label{eq:beta1,2}
\end{equation}
where
\begin{eqnarray}
&&\mathbf Q_{\alpha \alpha_1}(\mathbf q,t)=
\mathbf D_{\alpha \alpha_1}[\mathbf k (\mathbf q,t)]\exp\left(i\phi^\mathrm{(d)}_{\alpha\alpha_1}(\mathbf q,t)\right),
 \label{Q}
\\
&&\phi^\mathrm{(d)}_{\alpha\alpha_1}(\mathbf q,t)=
\nonumber
\\
 &&- \frac{1}{\hbar} \int_{-\infty}^t dt^\prime \left(E_\alpha[\mathbf k (\mathbf q,t^\prime)]-E_{\alpha_1}[\mathbf k (\mathbf q,t^\prime)]\right),
 \label{phi}
 \\ 
&&\mathbf D_{\alpha\alpha_1}=e \mathbfcal{A}_{\alpha\alpha_1}; ~
{\mathbfcal{A}}_{\alpha \alpha_1}({\mathbf q})=
\left\langle \Psi^{(\alpha)}_\mathbf q  |   i\frac{\partial}{\partial\mathbf q}|\Psi^{(\alpha_1)}_\mathbf q   \right\rangle .
\label{D}
\end{eqnarray} 
Here $ \mathrm{\Psi^{(\alpha)}_{{\bf k}}} $ are eigenfunctions of the Hamiltonian without an optical field, $H^\mathrm{TNN}+H^\mathrm{SOC}$,  where $\alpha \in \lbrace v,c_1,c_2 \rbrace$; matrix ${\mathbfcal A}_{\alpha \alpha_1}(\mathbf k)$ is non-Abelian Berry connection \cite{Wiczek_Zee_PhysRevLett.52_1984_Nonabelian_Berry_Phase, Xiao_Niu_RevModPhys.82_2010_Berry_Phase_in_Electronic_Properties, Yang_Liu_PhysRevB.90_2014_Non-Abelian_Berry_Curvature_and_Nonlinear_Optics}, $\mathbf D_{\alpha\alpha_1}$ is the interband dipole matrix, which determines optical transitions between the valence and conduction bands [see  Fig.\ 6 of appendix],
and $\phi^{\mathrm{(d)}}_{\alpha\alpha_1}(\mathbf q,t)$ is the dynamic phase. The trajectory in the reciprocal space, $\mathbf k(\mathbf q, t)$, is given by the Bloch theorem \cite{Bloch_Z_Phys_1929_Functions_Oscillations_in_Crystals}, $\mathbf k(\mathbf q, t)=\mathbf q-\frac{e}{\hbar}\int_{-\infty}^t \mathbf F(t^\prime) dt^\prime$; $\mathbf q$ is the initial crystal momentum.


 We numerically solve coupled ordinary differential Eqs.\ (\ref{eq:beta1,2} )  with initial conditions $\beta_{v\mathrm q}(-\infty)=1, \beta_{c_1\mathrm q}(-\infty)=0, \beta_{c_2\mathrm q}(-\infty)=0$. The total population of the CBs is calculated as
$N_{\mathrm{CB}}(\mathbf{q},t)=\left|\beta_{\mathrm c_1\mathbf{q}}(t)|^2+|\beta_{\mathrm c_2\mathbf{q}}(t)\right|^2$.
After the pulse ends, there remains residual CB population $N^\mathrm{(res)}_\mathrm{CB}(\mathbf{q})=N_{\mathrm{CB}}(\mathbf{q},\infty)$.

The field of a single-oscillation right-hand polarized pulse [see Eq.\ (\ref{Field})] is displayed in Fig.\ \ref{MoS2}(a) and the $\mathcal T$-reversed, left-hand pulse in Fig.\ \ref{MoS2}(b).
The residual CB population in the reciprocal space for $\mathrm{MoS_2}$ induced by such pulses with an amplitude of $0.2 ~\mathrm{V\AA^{-1}}$ is displayed  in  Fig. \ref{MoS2} for spin-up [$s_z=1/2$ or $\uparrow$, panels (c) and (d)] and spin-down [$s_z=-1/2$ or $\downarrow$, panels (e) and (f)]. The valley selectivity for either spin is very high: the left-handed pulse populates predominantly the $\mathrm K$ valleys, while the right-handed pulse mostly the $\mathrm{K^\prime}$ valleys. Such a difference in the populations of the $\mathrm K$ vs.\ $\mathrm{K^\prime}$ valleys is referred to as valley polarization. Additionally, for each handedness, there is a significant spin polarization (dependence of the population on spin). Protected by the $\mathcal T$-symmetry, the $\mathrm K_\uparrow$-valley population for a given handedness pulse is inversed ($\mathbf k\leftrightarrow-\mathbf k$) to the $\mathrm K^\prime_\downarrow$-valley population for the opposite handedness; the same is true for $\mathrm K_\downarrow$ and $\mathrm K^\prime_\uparrow$. Correspondingly, panel (c) is center-reflected to panel (f), and panel (d) to panel (e). 

We also performed computations for a two-oscillation pulse (see appendix Fig.\ 7) and found no fundamental difference from the single-oscillation pulse. In fact, both the valley polarization and CB population become higher.

The valley and spin polarization in TMDCs caused by circularly-polarized continuous-wave radiation \cite{Xiao_et_al_PRL_2012_Coupled_Spin_and_Valley_Physics, Zeng_et_al_Nature_Nanotechnology_2012_Valley_polarization_in_MoS2, Heinz_et_al_10.1038_Nnano.2012.96_Valley_Polarization_in_TMDC_by_Optical_Helicity} and 
relatively long 30 fs pulses \cite{Wang-2015-Ultrafast_Multi-Level_Logic_Gates}
were previousl known and attributed to angular momentum conservation at the $K, K^\prime$-points  \cite{Xiao_et_al_PRL_2012_Coupled_Spin_and_Valley_Physics, Feng_et_al_ncomms1882_2012_Valley_Selective_CD}. 
The spin polarization is related to the intrinsic SOC in the transition metals \cite{Xiao_et_al_PRL_2012_Coupled_Spin_and_Valley_Physics, Feng_et_al_ncomms1882_2012_Valley_Selective_CD, Liu_et_al_PRB_2014_Three_Band_Model}. In fact, SOC causes a significant spin splitting of the bands near the $\mathrm K$- and $\mathrm K^\prime$-points, which leads to different bandgaps in a given valley for the spin-up and spin-down bands
(see  Fig. 5 of appendix). 

A distinction of this work is that the significant CB-population and valley-polarization (along with  a smaller spin polarization) can be written by  a {\em single-oscillation} strong chiral pulse. The read-out can also be done by a single-oscillation chiral pulse: optical absorption of the read-out pulse of the same chirality will be reduced due to the Pauli blocking, while the opposite-chirality pulse absorption will not be attenuated because it interacts with the other, unpopulated valley. This one-optical-cycle recording and read-out make a basis of a fundamentally fastest optical memory.

\begin{figure}
\begin{center}\includegraphics[width=0.47\textwidth]{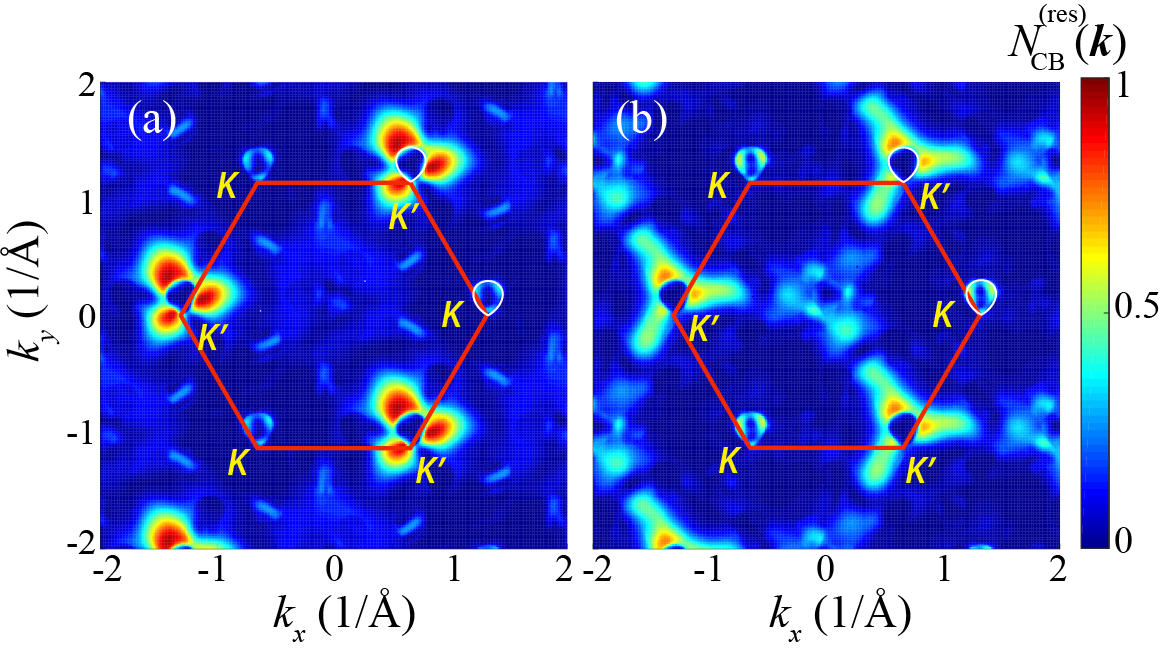}\end{center}
\caption{(Color online) Residual CB populations $N\mathrm{^{(res)}_\mathrm{CB}}(\mathbf{k})$ for monolayer $\mathrm{WS_2}$  after right-handed circularly polarized pulse. Note that the corresponding distributions for the right-handed pulses are related to these by the $\cal T$-symmetry similar to Fig.\ \ref{MoS2}. The red solid line shows the Brillouin zone boundary. Amplitude of the applied field is 0.2 $\mathrm{V\AA^{-1}}$. (a) Population $N\mathrm{^{(res)}_\mathrm{CB\downarrow}}(\mathbf{k})$ for  spin down electrons. (b) The same as panel (a) but for spin up electrons,  $N\mathrm{^{(res)}_\mathrm{CB\uparrow}}(\mathbf{k}).$ 
} 
\label{WS2}
\end{figure}

Figures \ref{WS2} (a) and (b) show the  residual CB population for another TMDC,  $\mathrm{WS_2}$, after a right-handed circularly polarized pulse with the amplitude of 0.2 $\mathrm {V\AA}^{-1}$  for spin up and spin down electrons, respectively. Similar to Fig.\ \ref{MoS2}, the right-handed pulse populates predominantly the vicinity of the $\mathrm K^\prime$ valleys in accord with the optical valley selection rule \cite{Liu_et_al_Chemical_Society_Rev_2015_Electronic_structures}. Due to stronger SOC in W in comparison to Mo, the spin dependence is even more pronounced in the distributions of  $N\mathrm{^{(res)}_{CB}}(\mathbf{k})$ in $\mathrm{WS_2}$ (Fig.\  \ref{WS2}) than in $\mathrm{MoS_2}$ (Fig.\ \ref{MoS2}).


\begin{figure}
\begin{center}\includegraphics[width=0.47\textwidth]{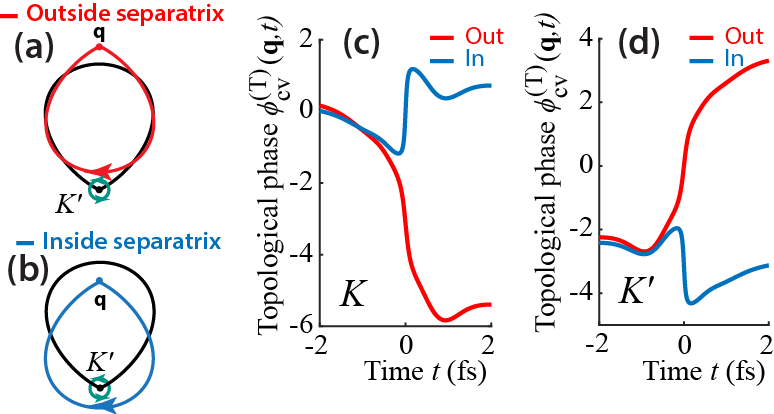}\end{center}
\caption{(Color online) 
For a chiral left-handed pulse, Bloch trajectories $\mathbf k(\mathbf q, t)$ in the $K^\prime$ valley and topological phase $\phi_\mathrm{cv}^\mathrm{(T)}(\mathbf q,t)$ for transitions $v\to c$ between bands forming the bandgaps at the $K$- and $K^\prime$-points. (a) Separatrix for the pulse used is shown by the black line. Electron Bloch trajectory $\mathbf k(\mathbf q,t)$ is shown for initial point $\mathbf q$ outside the separatrix. The $K^\prime$-point is denoted by a solid dot and the Berry connection (a counterpart of  vector potential) is denoted by a green ``whirl'' where the chirality is indicated by arrows. (b) The same as (a) but for  initial point $\mathbf q$ inside the separatrix. (c) Topological phase $\phi_\mathrm{cv}^\mathrm{(T)}(\mathbf q,t)$ on the Bloch trajectory for the $K$-point outside of the separatrix (red line) and inside the separatrix (blue line). (d) The same as (c) but for the $K^\prime$ point.
} 
\label{Phase_K-points}
\end{figure}

The valley selection rules for chiral pulses 
\cite{Niu_et_al_PhysRevB.77_2008_Valley_Selection_in_Graphene, Xiao_et_al_PRL_2012_Coupled_Spin_and_Valley_Physics, Zeng_et_al_Nature_Nanotechnology_2012_Valley_polarization_in_MoS2, Heinz_et_al_10.1038_Nnano.2012.96_Valley_Polarization_in_TMDC_by_Optical_Helicity, Feng_et_al_ncomms1882_2012_Valley_Selective_CD, Jones_et_al_Nature_Nanotechnology_2013_Optical_generation_of_excitonic_valley, Sie_et_al_Nature_Materials_2015_Valley-selective_optical} are related to the fact that the dipole moment in the plane and non-Abelian Berry connection are proportional -- see Eq.\ (\ref{D}).
These are angular momentum selection rules of linear optics, which are local in $\mathbf k$.

In contrast, there is also another, nonlinear-optical selection rule characteristic of strong-field excitation, which is evident from Figs.\ \ref{MoS2} and \ref{WS2}: In all cases when a given valley is favored by the angular momentum conservation, its population occurs {\em outside} of a closed curve (called separatrix \cite{Stockman_et_al_PhysRevB.93.155434_Graphene_Circular_Interferometry}). This is the case for $K^\prime$-valleys in Figs.\ \ref{MoS2} (c), (e) and Fig.\ \ref{WS2} and for the $K$-valleys in Fig.\ \ref{MoS2} (d), (f). In contrast, when the angular momentum selection rule suppresses a valley's population, then the momentum states {\em inside} the separatrix are more populated as is the case for the $K$-valleys in Figs.\ \ref{MoS2} (c), (e) and Fig.\ \ref{WS2}. 

\begin{figure}
\begin{center}\includegraphics[width=0.47\textwidth]{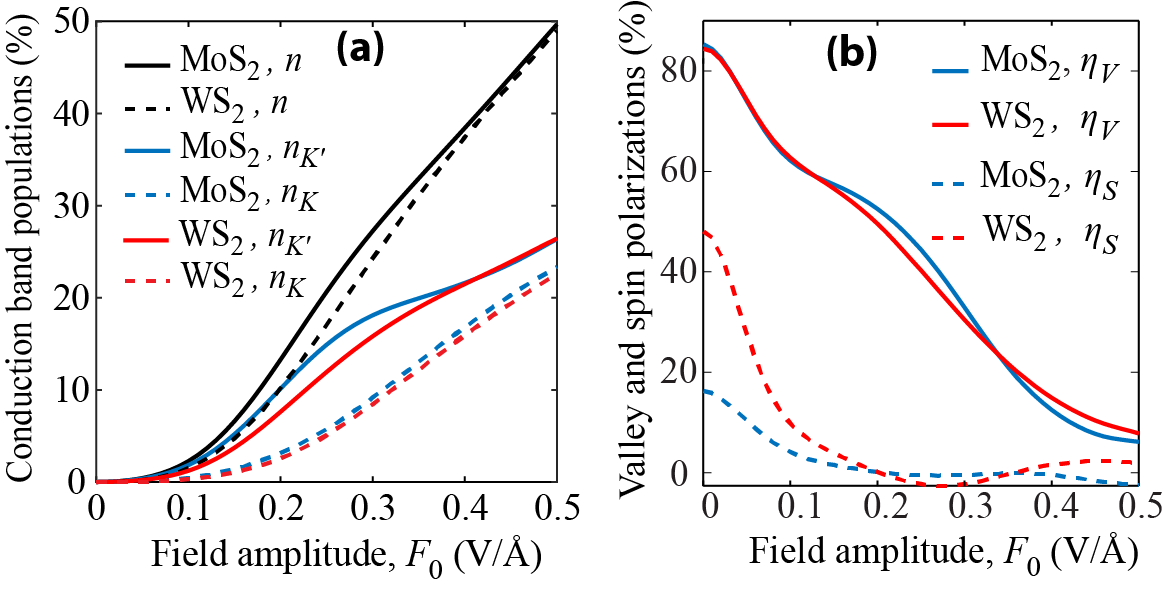}\end{center}\caption{(Color online) 
Valley CB populations, valley polarization, and spin polarization for TMDC $\mathrm{MoS_2}$ and $\mathrm{WS_2}$, as indicated. (a) Total CB's population $n$ and the CB's populations in the corresponding valleys as a function of the amplitude $F_0$ of the excitation right-handed pulse, color coded as indicated. (b) Same as in (a) but for valley and spin polarizations.
}
\label{Valley_Spin}
\end{figure}

Formation of such textures in the $K$ and $K^\prime$ valleys  is a fundamental effect directly related to the global topology of the Bloch bands. This effect is inherent in the strong-field excitation where an electron moves in the reciprocal space exploring the non-Abelian Berry connection, ${\mathbfcal A}_{\alpha \alpha_1}(\mathbf k)$, along its Bloch trajectory. We call it a {\em topological resonance}. It is also quantitatively important because it defines a fraction of the valley space favored for population and, consequently, the valley polarization (see below Fig.\ \ref{Valley_Spin} and its discussion). Also, such textures in the $k$-space cause electron currents in the real space, which we will consider elsewhere.

To understand the topological resonance, we turn to Figs.\ \ref{Phase_K-points} (a), (b). The separatrix, which is shown by a closed black line, is defined as a set of initial points $\mathbf q$ for which electron trajectories pass precisely through the corresponding $K$ or $K^\prime$ points \cite{Stockman_et_al_PhysRevB.93.155434_Graphene_Circular_Interferometry}. Its parametric equation is $\mathbf q(t)=\mathbf K-\mathbf k(0,t)$, or  $\mathbf q(t)=\mathbf K^\prime-\mathbf k(0,t)$ where $t\in (-\infty,\infty)$ is a parameter. Thus, the separatrix is an inverted electron trajectory starting at the $K$ or $K^\prime$ point.
 For initial crystal momentum $\mathbf q$ outside of the separatrix, the electron trajectory, $\mathbf k(\mathbf q,t)$, does not encircle the $K$-point as in Fig.\ \ref{Phase_K-points} (a), otherwise it does as in  Fig.\ \ref{Phase_K-points} (b). Because the coupling dipole matrix element is large at the $K$-points,
 the residual CB population will be enhanced close to the separatrix. 
 
 In Eq.\ (\ref{eq:beta1,2}), the interband coupling amplitude, $e\mathbf{F}(t)\mathbfcal A_{\alpha\alpha_1}[\mathbf k(\mathbf q,t)]$, acquires a nontrivial topological phase (non-Abelian Berry phase) $\phi^\mathrm{(T)}_{\alpha\alpha_1}[\mathbf k(\mathbf q,t)]=\arg\big\{\mathbf{F}(t)\mathbfcal{A}_{\alpha\alpha_1}[\mathbf k(\mathbf q,t)]\big\}$. This phase is displayed in Fig.\ \ref{Phase_K-points} (c)  for the $K$-valley  and in Fig.\ \ref{Phase_K-points} (d) for the $K^\prime$-valley. As we see, the changes of this phase for the valleys with opposite chiralities are opposite; for $\mathbf q$ outside of the separatrix, this change is significantly larger than otherwise (cf.\ the red vs. blue lines) and is close to $\pm 2\pi$.

The total phase, $\phi^{(\mathrm{tot})}_\mathrm{cv}(\mathbf q,t)=\arg\left\{\mathbf F(t)\mathbf Q_\mathrm{cv}\right\}$, of the interband coupling in Eq.\ (\ref{eq:beta1,2}) is a sum of the dynamic and topological phases,
\begin{equation}
 \phi^{(\mathrm{tot)}}_\mathrm{cv}(\mathbf q,t)=\phi^\mathrm{(d)}_\mathrm{cv}(\mathbf q,t)+\phi^\mathrm{(T)}_\mathrm{cv}(\mathbf q,t). 
 \label{phiQ}
 \end{equation}
 Note that the electron trajectory, $\mathbf k_T(\mathbf q,t)$, for the complete cycle is always closed. Thus the non-Abelian Berry phase, $\phi^\mathrm{(d)}_\mathrm{cv}(\mathbf q,t)$, is given by the integral of the non-Abelian Berry connection over a closed curve, which can be transformed to the integral of the non-Abelian Berry curvature (curl of the connection) over the area inside this curve. Consequently, the non-Abelian Berry phase for the entire pulse is gauge invariant.
 
 A significant ($\gtrsim2\pi$) change of $\phi^{(\mathrm{tot)}}_\mathrm{cv}(\mathbf q,t)$ along the Bloch trajectory, $\mathbf k(\mathbf q,t)$, leads to addition of amplitudes $\mathbf F(t) \mathbf Q_{\alpha\alpha_1}(\mathbf q, t) $ in Eq.\ (\ref{eq:beta1,2}) with varying signs, which tend to mutually annihilate each other. This prevents coherent accumulation of the CB population as defined by Eq.\ (\ref{eq:beta1,2}).  In contrast, the mutual cancellation of the dynamic and topological phases leads to the coherent (with the same phase) accumulation of the CB excitation amplitudes and enhanced CB population. This is the topological resonance effect. It is  absent if non-Abelian Berry connection $\mathbfcal A_{\alpha\alpha_1}(\mathbf q)=0$, in particular, for $\mathbf q$ in the vicinity of the $\Gamma$-point. Note that the a conventional resonance can also be described as cancellation between the dynamic phase $\Delta t/\hbar $ (where $\Delta$ is excitation energy) and the field phase $-\omega t$, which occurs for $\omega\approx\Delta/\hbar$.
 
Dynamic phase $\phi^\mathrm{(d)}_\mathrm{cv}(\mathbf q,t)$ [Eq.\ (\ref{phi})] monotonically decreases with time $t$ from 0 to $\approx -2\pi$. Hence,  the topological resonance takes place for the non-Abelian Berry phase, $\phi^\mathrm{(T)}_\mathrm{cv}(\mathbf q,t)$, increasing by $\approx 2\pi$. 
For a case of left-handed pulse illustrated in Fig.\ \ref{Phase_K-points}, the topological resonance occurs for crystal momentum $\mathbf q$ inside the separatrix for the $K$-point and outside of the separatrix for the $K^\prime$-point; in the latter case,  the CB population is also favored by the angular momentum selection rule  \cite{Xiao_et_al_PRL_2012_Coupled_Spin_and_Valley_Physics, Zeng_et_al_Nature_Nanotechnology_2012_Valley_polarization_in_MoS2, Heinz_et_al_10.1038_Nnano.2012.96_Valley_Polarization_in_TMDC_by_Optical_Helicity, Feng_et_al_ncomms1882_2012_Valley_Selective_CD, Jones_et_al_Nature_Nanotechnology_2013_Optical_generation_of_excitonic_valley, Sie_et_al_Nature_Materials_2015_Valley-selective_optical}. 
Protected by the $\mathcal T$-reversal symmetry, for the opposite chirality of the pulse, the $K$- and $K^\prime$-valleys are exchanged, and the spin is changed to the opposite -- cf.\ Fig.\ \ref{MoS2} and its discussion.

We quantify valley polarization as \\
$\eta_\mathrm V=({n^{\uparrow}_\mathrm{K^\prime}+n^{\downarrow}_\mathrm{K^\prime}-n^{\uparrow}_\mathrm{K}-n^{\downarrow}_\mathrm{K}})/({n^{\uparrow}_\mathrm{K^\prime}+n^{\downarrow}_\mathrm{K^\prime}+n^{\uparrow}_\mathrm{K}+n^{\downarrow}_\mathrm{K}})$,
where $n^{\uparrow}_\mathrm{K^\prime}$ is a CB population of the $K^\prime$-valley for spin-up electrons, and similar for other populations.
Likewise, we define spin polarization as \\
$\eta_S=({n^{\downarrow}_\mathrm{K'}-n^{\uparrow}_\mathrm{K'}+n^{\downarrow}_\mathrm{K}-n^{\uparrow}_\mathrm{K}})/({n^{\uparrow}_\mathrm{K'}+n^{\downarrow}_\mathrm{K'}+n^{\uparrow}_\mathrm{K}+n^{\downarrow}_\mathrm{K}})$.

Figure \ref{Valley_Spin} (a) displays total population $n=n_\mathrm{K}+n_\mathrm{K^\prime}$ and valley populations $n_{K}=n^{\uparrow}_{K}+n^{\downarrow}_{K}$, $n_{K^\prime}=n^{\uparrow}_{K^\prime}+n^{\downarrow}_{K^\prime}$ for monolayers of $\mathrm{MoS_2}$  and $\mathrm{WS_2}$  as functions of the amplitude, $F_0$, of a chiral left-handed excitation pulse. 
 As one can see, for both TMDC's, there is a strong asymmetry in the population: the $K'$ valley is preferentially populated. With an increase of $F_0$, this asymmetry decreases but the total CB population increases. 
 A reasonable compromise is $F_0\sim 0.2~\mathrm{V/\AA}$ where the CB population is high enough ($\sim 20\%$) but valley polarization is still also high: $\eta_\mathrm{V}\sim 60\%$. In contrast, the spin polarization is low,  $\eta_\mathrm{S}\sim 1\%$. 

To conclude, we have demonstrated a fundamental possibility to induce a significant CB populations and valley polarization in TMDCs during just one optical period of a chiral, moderately-high-field ($F_0\sim 0.2~\mathrm{V/\AA}$) laser pulse. This is a wide-band ultrafast process which is defined by a combination of the local angular-momentum conservation and the topological resonance that we have introduced. This resonance is due to mutual compensation of the dynamic phase and the non-Abelian Berry phase, which brings about formation of textures in the $\mathbf k$-space with discontinuities at the separatrices. These textures can be directly observed using the time-resolved angle-resolved photoelectron spectroscopy \cite{Chiang_et_al_PhysRevLett.107_Berry_Phase_in_Graphene_ARPES}.

The topological resonances can be present and pronounced not only in TMDCs but also in other materials with bandgaps at the Brillouin zone boundary, e.g., hexagonal boron nitride and others \cite{Xu_et_al_Chem_Rev_2013_Graphene-Like_2D_Materials} with bandgaps not at the $\Gamma$-point. The presence of the bandgap is essential because it causes a gradual accumulation of the non-Abelian Berry phase along the Bloch $\mathbf k$-space electron trajectory, which is necessary to compensate the gradually accumulating dynamic phase. In contrast, in gapless materials such as graphene, silicene, germanene, and surfaces of topological insulators, the non-Abelian Berry curvature has a $\delta$-function singularity at the Dirac points. Consequently, the non-Abelian Berry phase is accumulated discontinuously (at the Dirac points), and the topological resonances are not pronounced -- cf.\ Refs.\ \cite{Stockman_et_al_PhysRevB.93.155434_Graphene_Circular_Interferometry, Stockman_et_al_PhysRevB.96_2017_Berry_Phase}. 
The predicted induction of the valley polarization promises a wide range of important valleytronics applications to  PHz-band information processing and storage. The predicted topological resonance is a new concept, which promises novel developments in topological optics. In particular, the chiral, non-uniform electron distributions in the reciprocal space will cause chiral currents in the real space, which we will consider elsewhere.

\begin{acknowledgments}
Major funding was provided by Grant No. DE-FG02-01ER15213 from the Chemical Sciences, Biosciences and Geosciences Division, Office of Basic Energy Sciences, Office of Science, US Department of Energy. Numerical simulations have been performed using support by Grant No. DE-FG02-11ER46789 from the Materials Sciences and Engineering Division of the Office of the Basic Energy Sciences, Office of Science, U.S. Department of Energy. The work of V.A. was supported by Grant No. ECCS-1308473 from NSF. Support for S.A.O.M. came from a MURI Grant No. FA9550-15-1-0037 from the US Air Force of Scientific Research, and for J.-S.W. from a NSF EFRI NewLAW Grant EFMA-17 41691.
\end{acknowledgments}
\section{Appendix}

\section{Tight binding Hamiltonian}
The three band  third-nearest-neighbor (TNN) tight-binding (TB) model Hamiltonian, $H^\mathrm{{TNN}}$ , of a transition metal dichalcogenide (TMDC) monolayer is constructed from three orbitals (${d_{z^2}}$, ${d_{xy}}$, and ${d_{x^2-y^2}}$) of the metal atom, as introduced by Liu et al.\ \cite{Liu_et_al_PRB_2014_Three_Band_Model}, is
\begin{equation}
H^\mathrm{TNN}(\mathbf k)=\left[ {\begin{array}{ccc}
V_0 &V_1 & V_2\\
V^*_1 &V_{11} & V_{12}\\
V^*_2 &V^*_{12} & V_{22}\
\end{array} } \right]~,
\label{eq:Hamiltonian}
\end{equation}
where 
\begin{eqnarray}
{V_0}&=&{\epsilon_1}+{2t_0 (2 \cos\alpha \cos\beta+\cos2\alpha)}\nonumber\\&+&
2r_0(2\cos3\alpha \cos\beta +\cos2\beta)\nonumber\\&+&
{2u_0(2\cos2\alpha \cos2\beta +\cos4\alpha}),\nonumber\\
{\mathrm{Re}[V_1]}&=&-2 \sqrt{3}t_2\sin\alpha \sin\beta+
   {2(r_1+r_2)\sin3\alpha \sin\beta}\nonumber\\&-&
   {2\sqrt{3}u_2\sin2\alpha \sin2\beta},\nonumber\\
{\mathrm{Im}}[V_1]&=&2t_1\sin \alpha(2\cos\alpha +\cos \beta)\nonumber\\&+&
  2(r_1-r_2)  \sin3\alpha \cos\beta\nonumber\\&+&
  {2u_1\sin2\alpha (2\cos2\alpha+\cos2\beta)},\nonumber\\
{\mathrm{Re}}[V_2]&=&2t_2(\cos2 \alpha - \cos \alpha \cos \beta)\nonumber\\&-&
 \frac{2}{\sqrt{3}}(r_1+r_2)  (\cos3\alpha \cos\beta - \cos 2\beta)\nonumber\\&+&
 {2u_2(\cos4\alpha-\cos2\alpha \cos2\beta)},\nonumber\\
{\mathrm{Im}}[V_2]&=&2\sqrt{3}t_1\cos\alpha \sin\beta\nonumber\\&+&
 \frac{2}{\sqrt{3}}\sin\beta (r_1-r_2)(\cos3\alpha+2\cos\beta)\nonumber\\&+&
 { 2\sqrt{3}u_1\cos2\alpha \sin2\beta},\nonumber\\
{V_{11}}&=&\epsilon_2+(t_{11}+3t_{22})\cos\alpha \cos\beta\nonumber\\&+&2t_{11}\cos2\alpha+
  4r_{11}\cos3\alpha \cos\beta+2(r_{11}\nonumber\\&+&\sqrt{3}r_{12}\cos2\beta)+
  (u_{11}+3u_{22})\cos2\alpha \cos2\beta\nonumber\\&+&2u_{11}\cos4\alpha,\nonumber\\
{\mathrm{Re}}[V_{12}]&=&\sqrt{3}(t_{22}-t_{11})\sin\alpha \sin\beta +4r_{12}\sin3\alpha \sin \beta \nonumber\\&+&
 {\sqrt{3}(u_{22}-u_{11}\sin2\alpha \sin2\beta)},\nonumber\\
\mathrm{Im}[V_{12}]&=&4t_{12}\sin\alpha(\cos\alpha-cos\beta)\nonumber\\&+&4u_{12}\sin2\alpha (\cos2\alpha-\cos2\beta),\nonumber\\
{V_{22}}&=&\epsilon_2+(3t_{11}+t_{22})\cos\alpha \cos \beta \nonumber\\&+&2t_{22}\cos2\alpha { +2r_{11}(2\cos3\alpha \cos \beta +\cos 2\beta)}\nonumber\\&+&
 \frac{2}{\sqrt{3}}r_{12}(4\cos3\alpha \cos \beta - \cos 2\beta)\nonumber\\ &+&
(3u_{11}+u_{22})\cos2\alpha \cos2\beta +2u_{22}\cos 4\alpha\nonumber
\end{eqnarray}
in which
\begin{equation}
{(\alpha,\beta)=\left(\frac{1}{2}k_xa,\frac{\sqrt{3}}{2}k_ya\right)}~.
\end{equation}
The values of parameters for $\mathrm{MoS_2}$ and $\mathrm{WS_2}$ can be found in the Table\ \ref{T1}.\\


\begin{table}
\begin{center}
\begin{tabular}{|c|c|c|c|c|c|c|c| }
\hline
\multirow{3}{2.5em}{} &$a$ & $\mathrm{\epsilon_1}$ & $\mathrm{\epsilon_2}$ & $\mathrm{t_0}$ &$\mathrm{t_1}$ & $\mathrm{t_2}$& $\mathrm{t_{11}}$ \\
  
 & $\mathrm{t_{12}}$&$\mathrm{t_{22}}$ &$\mathrm{r_0}$ & $\mathrm{r_1}$&$\mathrm{r_2}$ & $\mathrm{r_{11}}$ & $\mathrm{r_{12}}$  \\
  
 &$\mathrm{u_0}$&$\mathrm{u_1}$&$\mathrm{u_{2}}$ & $\mathrm{u_{11}}$ & $\mathrm{u_{12}}$ &$\mathrm{u_{22}}$& $\mathrm{\lambda}$ \\
\hline
 
 \multirow{3}{2.5em}{$\mathrm{MoS_2}$} &3.190&0.683&1.707&-0.146&-0.114&0.506&0.085\\
 &0.162&0.073&0.060&-0.236&0.067&0.016&0.087\\
 &-0.038&0.046&0.001&0.266&-0.176&-0.150 &0.073\\
 \hline
  \multirow{3}{2.5em}{$\mathrm{WS_2}$} &3.191&0.717&1.916&-0.152&-0.097&0.590&0.047\\
& 0.178&0.016&0.069&-0.261&0.107&-0.003&0.109\\
& -0.054&0.045&0.002&0.325&-0.206&-0.163&0.211 \\
 \hline
\end{tabular}

\end{center}
\caption{Fitted parameters for three band TNN TB on the first-principles (FP) band structure in generalized-gradient approximation (GGA) case, lattice constant(a), and SOC paramrter ($\lambda$). All quantities are in unit eV except a which is in unit $\mathrm{\AA}$\cite{Liu_et_al_PRB_2014_Three_Band_Model}.}
\label{T1}
\end{table}

\section{SOC contribution to the Hamiltonian}
The contribution of the spin orbit coupling (SOC), ${H^\mathrm{SOC}}$, to the total Hamiltonian written in the basis of $\left\lbrace |d_{z^2},\uparrow\left>\right., |d_{xy},\uparrow\left>\right.,|d_{x^2-y^2},\uparrow\left>\right.,|d_{z^2},\downarrow\left>\right., |d_{xy},\downarrow\left>\right. \right.,$ $|d_{x^2-y^2},\downarrow\left>\right.  \left. \right\rbrace $  is the following matrix \cite{Winkler_Springer_Berlin_Heidelberg_2003_Acceleration_of_Electrons_in_a_Crystal_Lattice,  Liu_et_al_PRB_2014_Three_Band_Model} :
\begin{equation}
{H^\mathrm{SOC}}=\lambda{\bf L}.{\bf S}=\left[ {\begin{array}{cc}
\frac{\mathrm{\lambda}}{2}{L_z} & \mathrm{0} \\
\mathrm{0} & -\frac{\mathrm{\lambda}}{2}{L_z}
\end{array} } \right]
\label{Hsoc}
\end{equation}
where $\lambda$ is the SOC parameter, and $L_z$ is the $z$-component of the orbital angular momentum \cite{Liu_et_al_PRB_2014_Three_Band_Model},
\begin{equation}
{L_z=}\left[ {\begin{array}{ccc}
{0} & 0 & 0\\
0 & 0 & {2i}\\
0 & {-2i} & 0
\end{array} } \right].
\end{equation}                                                                                                                                                                             
Therefore, $ {H^\mathrm{SOC}}$ is 2$\times$2 block diagonal Hamiltonian  
where the nonzero upper block corresponds to spin up and the nonzero lower block corresponds to  spin down \cite{Liu_et_al_PRB_2014_Three_Band_Model}.
 

\begin{figure}
\begin{center}\includegraphics[width=0.47\textwidth]{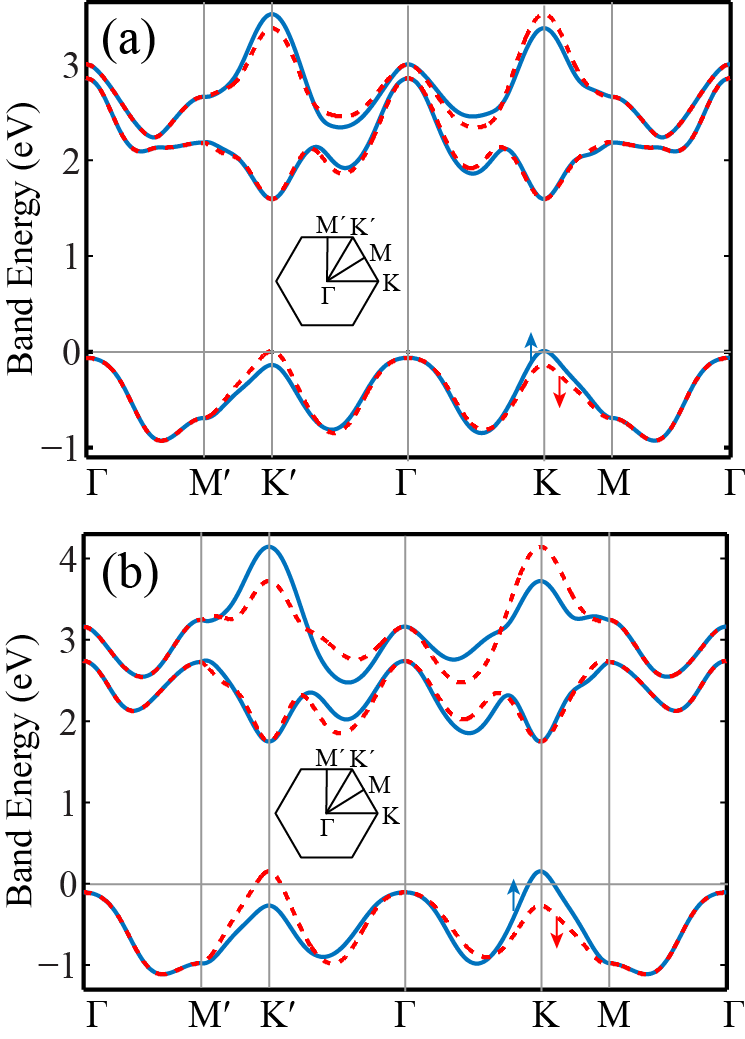}\end{center}
\caption{(Color online) Band structure for monolayers of (a) $\mathrm{MoS_2}$ and (b) $\mathrm{WS_2}$ for two component of the spin. The solid lines are for spin-up and the dash lines are for spin-down.
} 
\label{MoS2+WS2_SOC_TNN_G_G_G}
\end{figure}

\section{Main equations}
The total Hamiltonian, ${H_0(\mathbf{k})}$, in the same basis is 
\begin{equation}
{H_0(\mathbf{k})}=H^\mathrm{{TNN}}(\mathbf{k})+H^\mathrm{SOC}
\end{equation}
where $ {H^\mathrm{TNN}(\mathbf{k})} $ is the 3$ \times $3 tight binding Hamiltonian without spin, $ {H^\mathrm{SOC}(\mathbf{k})} $ is the SOC contribution, and the total Hamiltonian, $ {H_0(\mathbf k)} $, is a block diagonal operator expressed as   
\begin{eqnarray}
{H_0(\mathbf{k})}&=&\left[ {\begin{array}{cc}
{H^\mathrm{TNN}(\mathbf{k})+\frac{\mathrm{\lambda}}{2}{L_z}}&0\\0&{H^\mathrm{TNN}(\mathbf{k})-\frac{\mathrm{\lambda}}{2}{L_z}}
\end{array} } \right] \nonumber\\&=&\left[ {\begin{array}{cc}
{H^\mathrm{\uparrow}_{3\times3}(\mathbf{k})}&0\\0&{H^\mathrm{\downarrow}_{3\times3}(\mathbf{k})}
\end{array} } \right]~,
\label{eq:Total Hamiltonian}
\end{eqnarray}
in which the nonzero upper block corresponds to the spin up and the nonzero lower block to the spin down. Band structures of $ \mathrm{MoS_2} $ and $ \mathrm{WS_2} $  for the two components of the spin are shown in Fig. \ref{MoS2+WS2_SOC_TNN_G_G_G}, which shows spin splitting of the energy bands due to the intrinsic SOC. Protected by the time-reversal ($\mathcal T$) symmetry, the band energies in the $K$- and $K^\prime$-valleys are identical but the spins are reversed as illustrated in Fig. \ref{MoS2+WS2_SOC_TNN_G_G_G}.

In the presence of an external field, $\mathbf F(t)$, the Hamiltonian in the length gauge is $H_0(\mathbf k)+H^\mathrm{int}$, where $H^\mathrm{int}=e\mathbf F(t)\mathbf r$, and $e$ is unit charge. 
Electron dynamics in the presence of field ${\bf F}(t)$ includes two major components: intraband and  interband. The intraband electron dynamics in a single band is described by the Bloch acceleration theorem, ${{\bf{k}}({\bf{q}},t)={\bf{q}}+\frac{e}{\hbar c}{\bf{A}}(t)}$, where $\mathbf k(\mathbf q,t)$ is electron crystal momentum as a function of time $t$, $\mathbf q$ is the initial crystal momentum, ${{\mathbf{A}}(t)=-c\int^t_{-\infty}{{\mathbf{F}}(t')dt'}}$ is vector potential in the velocity gauge, and $c$ is speed of light. 

 We describe the resulting electron dynamics by solving time-dependent Schr\"odinger equation (TDSE). Since the Hamiltonian, $H_0(\mathbf k)$, is block-diagonal, the spin-up and spin-down components  are decoupled. Therefore,  the TDSE for each component of the spin,
\begin{equation}{
i \hbar \frac{d \Psi}{dt}= (H^{\mathrm{s}}_{3\times3}+H^{\mathrm{int}}) \Psi}~,
\end{equation}
where $ \mathrm{s}\in\left\{\uparrow, \downarrow\right\}$, can be solved independently.

\begin{figure}
\begin{center}\includegraphics[width=0.47\textwidth]{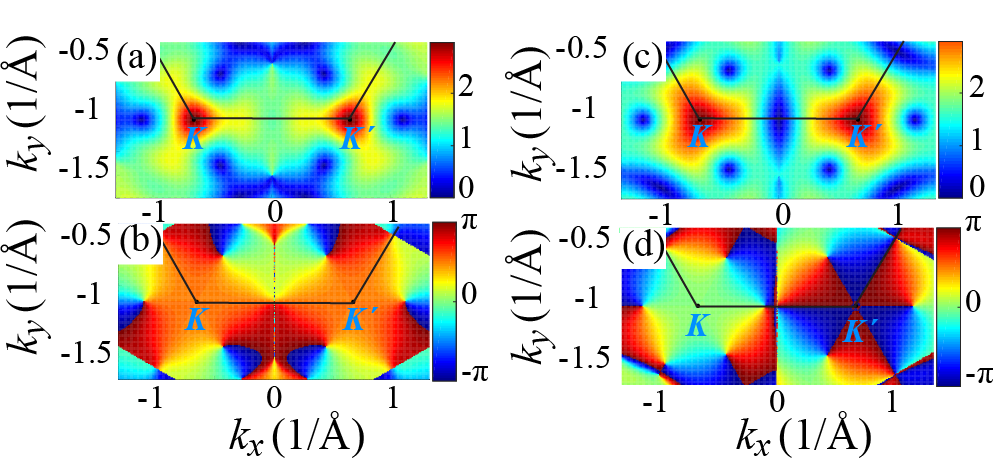}\end{center}
\caption{(Color online) Coupling dipole matrix element $\mathbf D$ for $\mathrm{MoS_2}$. (a)  Modulus of longitudinal component, $D_k=\mathbf{D}\hat{k}$, (b)  Phase of $D_k$, (c)  Modulus of tangential component ${D}_\varphi=\mathbf{D}\hat{\varphi}$, and (d)  Phase of tangential component ${D}_\varphi$  calculated in the vicinity of each valley. Black solid lines show the boundary of the Brillioun zone of $\mathrm{MoS_2}$.
} 
\label{MoS2_dipole_k_phi}
\end{figure}

In a single band, the solutions for the TDSE are Houston functions \cite{Houston_PR_1940_Electron_Acceleration_in_Lattice} 
\begin{equation}{
\Phi^{(H)}_{\alpha {\bf q}}({\bf r},t)=\Psi^{(\alpha)}_{{\bf k}({\bf q},t)}({\bf{r}})e^{{- \frac{i}{\hbar}} {\int} {dtE_\alpha[{\bf k }({\bf q},t)]}}}~,
\end{equation}
where $ {\Psi^{(\alpha)}_{{\mathbf k}}} $ are the eigenfunctions of ${H^\mathrm s_{3\times 3}}$,  $ {\alpha \in \lbrace v,c_1,c_2 \rbrace} $, and $ {v,c_1,c_2} $ are the highest valence band and two lowest conduction bands, respectively.
Using the set of the Houston functions as a basis, a general solution of the TDSE is expanded in the form
 \begin{equation}
 \Psi_{\bf q}({\bf r},t)=\sum_{\alpha=v,c_1,c_2}\beta_{\alpha{\bf q}}(t)\Phi^{(H)}_{{ \alpha}{\bf q}}({\bf r},t)~,
\end{equation}   
where $ \beta_{\alpha{\bf q}} $ are the expansion coefficients satisfying the following coupled ordinary differential equations,
\begin{eqnarray}
\frac{{d{\beta _{c_1{\bf{q}}}}(t)}}{{dt}}&=& 
- \frac{i}{\hbar}\mathbf F(t) \mathbf Q_{c_1 v}(\mathbf q,t) \beta _{v \mathbf q}(t)\nonumber\\ 
&-& \frac{i}{\hbar}\mathbf F(t) \mathbf Q_{c_1 c_2}(\mathbf q,t) \beta _{c_2 \mathbf q}(t) ~,
\label{eq:beta_c1}
\\
\frac{{d{\beta _{c_2{\bf{q}}}}(t)}}{{dt}} &=& 
- \frac{i}{\hbar}\mathbf F(t) \mathbf Q_{c_2 v}(\mathbf q,t) \beta _{v \mathbf q}(t) 
\nonumber\\&-& \frac{i}{\hbar}\mathbf F(t) \mathbf Q_{c_1 c_2}^\ast(\mathbf q,t) \beta _{c_1 \mathbf q}(t) ~,
\label{eq:beta_c2}
\\
\frac{{d{\beta _{v{\bf{q}}}}(t)}}{{dt}} &=&  
- \frac{i}{\hbar}\mathbf F(t) \mathbf Q_{c_1 v}^\ast(\mathbf q,t) \beta _{c_1 \mathbf q}(t) 
\nonumber\\&-& \frac{i}{\hbar}\mathbf F(t) \mathbf Q_{c_2 v}^\ast(\mathbf q,t) \beta _{c_2 \mathbf q}(t) ~,
\label{eq:beta_v}
\end{eqnarray}
and
\begin{eqnarray}
{\bf Q}_{\alpha \alpha_1}({\bf q},t)&=&e
\mathbfcal A_{\alpha \alpha_1}(\mathbf q,t)\exp\left(i\phi^\mathrm{(d)}_{\alpha\alpha_1}(\bf q,t)\right),\\
{\mathbfcal{A}}_{\alpha \alpha_1}({\mathbf q})&=& \left\langle \Psi^{(\alpha)}_{\bf {q}}| i\frac{\partial}{\partial\bf {q}}|\Psi^{(\alpha_1)}_{\bf {q}} \right\rangle,\\
 \phi^\mathrm{(d)}_{\alpha\alpha_1}(\mathbf q,t)&=&
- \frac{1}{\hbar} \int_{-\infty}^t dt^\prime \left(E_\alpha[\mathbf k (\mathbf q,t^\prime)]\right.\nonumber\\&-&\left.E_{\alpha_1}[\mathbf k (\mathbf q,t^\prime)]\right),
\end{eqnarray}
Here ${\mathbfcal{A}}_{\alpha \alpha_1}({\mathbf q})$ is non-Abelian Berry connection \cite{Wiczek_Zee_PhysRevLett.52_1984_Nonabelian_Berry_Phase, Xiao_Niu_RevModPhys.82_2010_Berry_Phase_in_Electronic_Properties, Yang_Liu_PhysRevB.90_2014_Non-Abelian_Berry_Curvature_and_Nonlinear_Optics},
 and $ \phi^\mathrm{(d)}_{\alpha\alpha_1}(\mathbf q,t)$ is the dynamic phase.
The interband dipole matrix, $\mathbf D_{\alpha\alpha_1}$ is simply related to the non-Abelian Berry connection, $\mathbf D_{\alpha\alpha_1}=e\mathbfcal A_{\alpha\alpha_1}$; it determines optical transitions between the valence and conduction bands. Figure \ref{MoS2_dipole_k_phi} shows the modulus and the phase of $\mathbf{D}$ for the longitudinal component in panels (a) and (b) and for the tangential component in panels (c) and (d) respectively. 

We solve the set of coupled ordinary differential equations (\ref{eq:beta_c1})-(\ref{eq:beta_v}) numerically by using  a variable time step Runge-Kutta method \cite{butcher_Cambridge_University_Press_1963_Runge_Kutta_integration_processes} with the following initial conditions ($ {\beta_{v \bf q},\beta_{c_1 \bf q}, \beta_{c_2 \bf q}  }$)=(1,0,0) to find the bands populations $N$ as a function of time and the lattice momentum $\mathbf{q}$.


\section{Mean Frequecny of the Optical Pulse}
An ultrafast pulse has no definite frequency since its Fourier component is widely distributed in the frequency space. We calculate the mean frequency, $\bar\omega$, of an optical pulse as 
\begin{equation}
\bar{\omega} = \frac{\int \omega S(\omega)d\omega}{\int  S(\omega)d\omega}~,
\end{equation}
where the pulse spectrum, $S(\omega)$, is defined as
\begin{equation}
S(\omega) = |\mathbf{F}_\omega|^2,~~\mathbf{F}_\omega=\int_{-\infty}^\infty\mathbf F(t)e^{i\omega t}dt~.
\end{equation}
For the pulse, described in Eq. (1) of the main text, we have calculated
$\hbar\bar{\omega}\simeq1.2$ eV.

 \section{Two-Oscillation Pulse}

\begin{figure}
\begin{center}\includegraphics[width=0.4\textwidth]{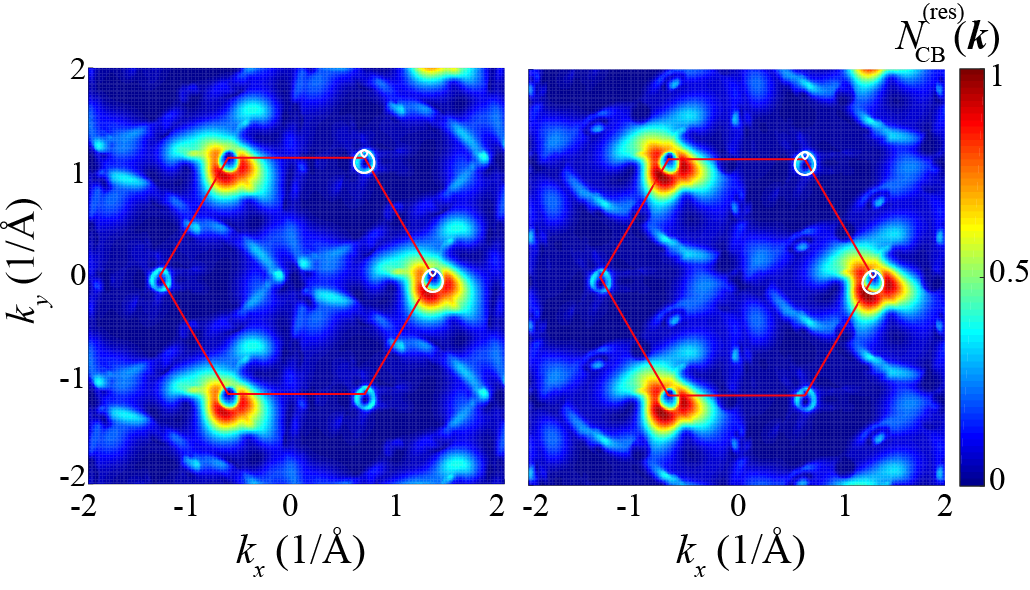}\end{center}
\caption{(Color online)  Residual CB populations $N\mathrm{^{(res)}_{\mathrm{CB},s}}(\mathbf{k})$ for monolayer $\mathrm{MoS_2}$  after left-handed circularly polarized pulse with two oscillations, see Eq.\ (\ref{Fxy}). The red solid line shows the Brillouin zone boundary. Amplitude of the applied field is  $F_0=0.2~~\mathrm{V\AA^{-1}}$. (a) Population $N\mathrm{^{(res)}_\mathrm{CB\uparrow}}(\mathbf{k})$ for  spin up electrons. (b) The same as panel (a) but for spin down electrons,  $N\mathrm{^{(res)}_\mathrm{CB\downarrow}}(\mathbf{k}).$ 
} 
\label{Spin_up_down_MoS2_H3_H4}
\end{figure}

Here we provide a solution for a pulse which is longer than the pulse used in the main text. This pulse contains two optical field oscillations of the same chirality and is parametrized as
\begin{eqnarray}
&& F_x(t)=F_0\frac{1}{12}e^{-u^2}H^{(4)}(u)~,
\nonumber
\\
&&  F_y(t)=F_0\frac{1}{4}e^{-u^2}H^{(3)}(u)~,
\label{Fxy}
\end{eqnarray}
where $u=t/\tau$, and $H^{(n)}$ is a Hermite polynomial of power $n$.

Calculated spin-resolved population distributions in the CB of MoS$_2$ for a two-oscillation left-handed pulse of Eq.\ (\ref{Fxy}) is displayed in Fig.\ \ref{Spin_up_down_MoS2_H3_H4}. Comparing it to Figs.\ 2 (c), (e) of the main text, one can conclude that there is no qualitative changes in CB population distribution when extra oscillations are added to the pulse. There are some changes of the distributions along the separatrix but general picture remains the same. Namely, the $K$-valleys are predominantly populated outside of the separatrix. There is a very small population of the $K^\prime$-valleys inside the separatrix. Overall, the valley polarization is higher that for a single-oscillation pulse. This is understandable because the two-oscillation pulse is closer to a circularly polarized CW radiation that the single oscillation one.



\begin{thebibliography}{62}%
\makeatletter
\providecommand \@ifxundefined [1]{%
 \@ifx{#1\undefined}
}%
\providecommand \@ifnum [1]{%
 \ifnum #1\expandafter \@firstoftwo
 \else \expandafter \@secondoftwo
 \fi
}%
\providecommand \@ifx [1]{%
 \ifx #1\expandafter \@firstoftwo
 \else \expandafter \@secondoftwo
 \fi
}%
\providecommand \natexlab [1]{#1}%
\providecommand \enquote  [1]{``#1''}%
\providecommand \bibnamefont  [1]{#1}%
\providecommand \bibfnamefont [1]{#1}%
\providecommand \citenamefont [1]{#1}%
\providecommand \href@noop [0]{\@secondoftwo}%
\providecommand \href [0]{\begingroup \@sanitize@url \@href}%
\providecommand \@href[1]{\@@startlink{#1}\@@href}%
\providecommand \@@href[1]{\endgroup#1\@@endlink}%
\providecommand \@sanitize@url [0]{\catcode `\\12\catcode `\$12\catcode
  `\&12\catcode `\#12\catcode `\^12\catcode `\_12\catcode `\%12\relax}%
\providecommand \@@startlink[1]{}%
\providecommand \@@endlink[0]{}%
\providecommand \url  [0]{\begingroup\@sanitize@url \@url }%
\providecommand \@url [1]{\endgroup\@href {#1}{\urlprefix }}%
\providecommand \urlprefix  [0]{URL }%
\providecommand \Eprint [0]{\href }%
\providecommand \doibase [0]{http://dx.doi.org/}%
\providecommand \selectlanguage [0]{\@gobble}%
\providecommand \bibinfo  [0]{\@secondoftwo}%
\providecommand \bibfield  [0]{\@secondoftwo}%
\providecommand \translation [1]{[#1]}%
\providecommand \BibitemOpen [0]{}%
\providecommand \bibitemStop [0]{}%
\providecommand \bibitemNoStop [0]{.\EOS\space}%
\providecommand \EOS [0]{\spacefactor3000\relax}%
\providecommand \BibitemShut  [1]{\csname bibitem#1\endcsname}%
\let\auto@bib@innerbib\@empty
\bibitem [{\citenamefont {Fattahi}\ \emph {et~al.}(2014)\citenamefont
  {Fattahi}, \citenamefont {Barros}, \citenamefont {Gorjan}, \citenamefont
  {Nubbemeyer}, \citenamefont {Alsaif}, \citenamefont {Teisset}, \citenamefont
  {Schultze}, \citenamefont {Prinz}, \citenamefont {Haefner}, \citenamefont
  {Ueffing}, \citenamefont {Alismail}, \citenamefont {Vamos}, \citenamefont
  {Schwarz}, \citenamefont {Pronin}, \citenamefont {Brons}, \citenamefont
  {Geng}, \citenamefont {Arisholm}, \citenamefont {Ciappina}, \citenamefont
  {Yakovlev}, \citenamefont {Kim}, \citenamefont {Azzeer}, \citenamefont
  {Karpowicz}, \citenamefont {Sutter}, \citenamefont {Major}, \citenamefont
  {Metzger},\ and\ \citenamefont
  {Krausz}}]{Fattahi_Third-generation_femtosecond_technology_Optica_2014}%
  \BibitemOpen
  \bibfield  {author} {\bibinfo {author} {\bibfnamefont {H.}~\bibnamefont
  {Fattahi}}, \bibinfo {author} {\bibfnamefont {H.~G.}\ \bibnamefont {Barros}},
  \bibinfo {author} {\bibfnamefont {M.}~\bibnamefont {Gorjan}}, \bibinfo
  {author} {\bibfnamefont {T.}~\bibnamefont {Nubbemeyer}}, \bibinfo {author}
  {\bibfnamefont {B.}~\bibnamefont {Alsaif}}, \bibinfo {author} {\bibfnamefont
  {C.~Y.}\ \bibnamefont {Teisset}}, \bibinfo {author} {\bibfnamefont
  {M.}~\bibnamefont {Schultze}}, \bibinfo {author} {\bibfnamefont
  {S.}~\bibnamefont {Prinz}}, \bibinfo {author} {\bibfnamefont
  {M.}~\bibnamefont {Haefner}}, \bibinfo {author} {\bibfnamefont
  {M.}~\bibnamefont {Ueffing}}, \bibinfo {author} {\bibfnamefont
  {A.}~\bibnamefont {Alismail}}, \bibinfo {author} {\bibfnamefont
  {L.}~\bibnamefont {Vamos}}, \bibinfo {author} {\bibfnamefont
  {A.}~\bibnamefont {Schwarz}}, \bibinfo {author} {\bibfnamefont
  {O.}~\bibnamefont {Pronin}}, \bibinfo {author} {\bibfnamefont
  {J.}~\bibnamefont {Brons}}, \bibinfo {author} {\bibfnamefont {X.~T.}\
  \bibnamefont {Geng}}, \bibinfo {author} {\bibfnamefont {G.}~\bibnamefont
  {Arisholm}}, \bibinfo {author} {\bibfnamefont {M.}~\bibnamefont {Ciappina}},
  \bibinfo {author} {\bibfnamefont {V.~S.}\ \bibnamefont {Yakovlev}}, \bibinfo
  {author} {\bibfnamefont {D.-E.}\ \bibnamefont {Kim}}, \bibinfo {author}
  {\bibfnamefont {A.~M.}\ \bibnamefont {Azzeer}}, \bibinfo {author}
  {\bibfnamefont {N.}~\bibnamefont {Karpowicz}}, \bibinfo {author}
  {\bibfnamefont {D.}~\bibnamefont {Sutter}}, \bibinfo {author} {\bibfnamefont
  {Z.}~\bibnamefont {Major}}, \bibinfo {author} {\bibfnamefont
  {T.}~\bibnamefont {Metzger}}, \ and\ \bibinfo {author} {\bibfnamefont
  {F.}~\bibnamefont {Krausz}},\ }\href@noop {} {\bibfield  {journal} {\bibinfo
  {journal} {Optica}\ }\textbf {\bibinfo {volume} {1}},\ \bibinfo {pages} {45}
  (\bibinfo {year} {2014})}\BibitemShut {NoStop}%
\bibitem [{\citenamefont {Schiffrin}\ \emph {et~al.}(2012)\citenamefont
  {Schiffrin}, \citenamefont {Paasch-Colberg}, \citenamefont {Karpowicz},
  \citenamefont {Apalkov}, \citenamefont {Gerster}, \citenamefont {Muhlbrandt},
  \citenamefont {Korbman}, \citenamefont {Reichert}, \citenamefont {Schultze},
  \citenamefont {Holzner}, \citenamefont {Barth}, \citenamefont {Kienberger},
  \citenamefont {Ernstorfer}, \citenamefont {Yakovlev}, \citenamefont
  {Stockman},\ and\ \citenamefont
  {Krausz}}]{Schiffrin_at_al_Nature_2012_Current_in_Dielectric}%
  \BibitemOpen
  \bibfield  {author} {\bibinfo {author} {\bibfnamefont {A.}~\bibnamefont
  {Schiffrin}}, \bibinfo {author} {\bibfnamefont {T.}~\bibnamefont
  {Paasch-Colberg}}, \bibinfo {author} {\bibfnamefont {N.}~\bibnamefont
  {Karpowicz}}, \bibinfo {author} {\bibfnamefont {V.}~\bibnamefont {Apalkov}},
  \bibinfo {author} {\bibfnamefont {D.}~\bibnamefont {Gerster}}, \bibinfo
  {author} {\bibfnamefont {S.}~\bibnamefont {Muhlbrandt}}, \bibinfo {author}
  {\bibfnamefont {M.}~\bibnamefont {Korbman}}, \bibinfo {author} {\bibfnamefont
  {J.}~\bibnamefont {Reichert}}, \bibinfo {author} {\bibfnamefont
  {M.}~\bibnamefont {Schultze}}, \bibinfo {author} {\bibfnamefont
  {S.}~\bibnamefont {Holzner}}, \bibinfo {author} {\bibfnamefont {J.~V.}\
  \bibnamefont {Barth}}, \bibinfo {author} {\bibfnamefont {R.}~\bibnamefont
  {Kienberger}}, \bibinfo {author} {\bibfnamefont {R.}~\bibnamefont
  {Ernstorfer}}, \bibinfo {author} {\bibfnamefont {V.~S.}\ \bibnamefont
  {Yakovlev}}, \bibinfo {author} {\bibfnamefont {M.~I.}\ \bibnamefont
  {Stockman}}, \ and\ \bibinfo {author} {\bibfnamefont {F.}~\bibnamefont
  {Krausz}},\ }\href@noop {} {\bibfield  {journal} {\bibinfo  {journal}
  {Nature}\ }\textbf {\bibinfo {volume} {493}},\ \bibinfo {pages} {70}
  (\bibinfo {year} {2012})}\BibitemShut {NoStop}%
\bibitem [{\citenamefont {Schultze}\ \emph {et~al.}(2013)\citenamefont
  {Schultze}, \citenamefont {Bothschafter}, \citenamefont {Sommer},
  \citenamefont {Holzner}, \citenamefont {Schweinberger}, \citenamefont
  {Fiess}, \citenamefont {Hofstetter}, \citenamefont {Kienberger},
  \citenamefont {Apalkov}, \citenamefont {Yakovlev}, \citenamefont {Stockman},\
  and\ \citenamefont
  {Krausz}}]{Schultze_et_al_Nature_2012_Controlling_Dielectrics}%
  \BibitemOpen
  \bibfield  {author} {\bibinfo {author} {\bibfnamefont {M.}~\bibnamefont
  {Schultze}}, \bibinfo {author} {\bibfnamefont {E.~M.}\ \bibnamefont
  {Bothschafter}}, \bibinfo {author} {\bibfnamefont {A.}~\bibnamefont
  {Sommer}}, \bibinfo {author} {\bibfnamefont {S.}~\bibnamefont {Holzner}},
  \bibinfo {author} {\bibfnamefont {W.}~\bibnamefont {Schweinberger}}, \bibinfo
  {author} {\bibfnamefont {M.}~\bibnamefont {Fiess}}, \bibinfo {author}
  {\bibfnamefont {M.}~\bibnamefont {Hofstetter}}, \bibinfo {author}
  {\bibfnamefont {R.}~\bibnamefont {Kienberger}}, \bibinfo {author}
  {\bibfnamefont {V.}~\bibnamefont {Apalkov}}, \bibinfo {author} {\bibfnamefont
  {V.~S.}\ \bibnamefont {Yakovlev}}, \bibinfo {author} {\bibfnamefont {M.~I.}\
  \bibnamefont {Stockman}}, \ and\ \bibinfo {author} {\bibfnamefont
  {F.}~\bibnamefont {Krausz}},\ }\href@noop {} {\bibfield  {journal} {\bibinfo
  {journal} {Nature}\ }\textbf {\bibinfo {volume} {493}},\ \bibinfo {pages}
  {75} (\bibinfo {year} {2013})}\BibitemShut {NoStop}%
\bibitem [{\citenamefont {Mitrofanov}\ \emph {et~al.}(2011)\citenamefont
  {Mitrofanov}, \citenamefont {Verhoef}, \citenamefont {Serebryannikov},
  \citenamefont {Lumeau}, \citenamefont {Glebov}, \citenamefont {Zheltikov},\
  and\ \citenamefont {{Baltu\v
  ska}}}]{Baltuska_et_al_Attosecond_IonizationPRL_2011}%
  \BibitemOpen
  \bibfield  {author} {\bibinfo {author} {\bibfnamefont {A.~V.}\ \bibnamefont
  {Mitrofanov}}, \bibinfo {author} {\bibfnamefont {A.~J.}\ \bibnamefont
  {Verhoef}}, \bibinfo {author} {\bibfnamefont {E.~E.}\ \bibnamefont
  {Serebryannikov}}, \bibinfo {author} {\bibfnamefont {J.}~\bibnamefont
  {Lumeau}}, \bibinfo {author} {\bibfnamefont {L.}~\bibnamefont {Glebov}},
  \bibinfo {author} {\bibfnamefont {A.~M.}\ \bibnamefont {Zheltikov}}, \ and\
  \bibinfo {author} {\bibfnamefont {A.}~\bibnamefont {{Baltu\v ska}}},\
  }\href@noop {} {\bibfield  {journal} {\bibinfo  {journal} {Phys. Rev. Lett.}\
  }\textbf {\bibinfo {volume} {106}},\ \bibinfo {pages} {147401} (\bibinfo
  {year} {2011})}\BibitemShut {NoStop}%
\bibitem [{\citenamefont {Goulielmakis}\ \emph {et~al.}(2007)\citenamefont
  {Goulielmakis}, \citenamefont {Yakovlev}, \citenamefont {Cavalieri},
  \citenamefont {Uiberacker}, \citenamefont {Pervak}, \citenamefont
  {Apolonski}, \citenamefont {Kienberger}, \citenamefont {Kleineberg},\ and\
  \citenamefont
  {Krausz}}]{Goulielmakis_et_al_Lightwave_Electronics_Science_2007}%
  \BibitemOpen
  \bibfield  {author} {\bibinfo {author} {\bibfnamefont {E.}~\bibnamefont
  {Goulielmakis}}, \bibinfo {author} {\bibfnamefont {V.~S.}\ \bibnamefont
  {Yakovlev}}, \bibinfo {author} {\bibfnamefont {A.~L.}\ \bibnamefont
  {Cavalieri}}, \bibinfo {author} {\bibfnamefont {M.}~\bibnamefont
  {Uiberacker}}, \bibinfo {author} {\bibfnamefont {V.}~\bibnamefont {Pervak}},
  \bibinfo {author} {\bibfnamefont {A.}~\bibnamefont {Apolonski}}, \bibinfo
  {author} {\bibfnamefont {R.}~\bibnamefont {Kienberger}}, \bibinfo {author}
  {\bibfnamefont {U.}~\bibnamefont {Kleineberg}}, \ and\ \bibinfo {author}
  {\bibfnamefont {F.}~\bibnamefont {Krausz}},\ }\href@noop {} {\bibfield
  {journal} {\bibinfo  {journal} {Science}\ }\textbf {\bibinfo {volume}
  {317}},\ \bibinfo {pages} {769} (\bibinfo {year} {2007})}\BibitemShut
  {NoStop}%
\bibitem [{\citenamefont {Paasch-Colberg}\ \emph {et~al.}(2016)\citenamefont
  {Paasch-Colberg}, \citenamefont {Kruchinin}, \citenamefont {Saglam},
  \citenamefont {Kapser}, \citenamefont {Cabrini}, \citenamefont {Muehlbrandt},
  \citenamefont {Reichert}, \citenamefont {Barth}, \citenamefont {Ernstorfer},
  \citenamefont {Kienberger}, \citenamefont {Yakovlev}, \citenamefont
  {Karpowicz},\ and\ \citenamefont
  {Schiffrin}}]{Paasch_Colberg_et_al_Optica_2016_optical_control}%
  \BibitemOpen
  \bibfield  {author} {\bibinfo {author} {\bibfnamefont {T.}~\bibnamefont
  {Paasch-Colberg}}, \bibinfo {author} {\bibfnamefont {S.~Y.}\ \bibnamefont
  {Kruchinin}}, \bibinfo {author} {\bibfnamefont {O.}~\bibnamefont {Saglam}},
  \bibinfo {author} {\bibfnamefont {S.}~\bibnamefont {Kapser}}, \bibinfo
  {author} {\bibfnamefont {S.}~\bibnamefont {Cabrini}}, \bibinfo {author}
  {\bibfnamefont {S.}~\bibnamefont {Muehlbrandt}}, \bibinfo {author}
  {\bibfnamefont {J.}~\bibnamefont {Reichert}}, \bibinfo {author}
  {\bibfnamefont {J.~V.}\ \bibnamefont {Barth}}, \bibinfo {author}
  {\bibfnamefont {R.}~\bibnamefont {Ernstorfer}}, \bibinfo {author}
  {\bibfnamefont {R.}~\bibnamefont {Kienberger}}, \bibinfo {author}
  {\bibfnamefont {V.~S.}\ \bibnamefont {Yakovlev}}, \bibinfo {author}
  {\bibfnamefont {N.}~\bibnamefont {Karpowicz}}, \ and\ \bibinfo {author}
  {\bibfnamefont {A.}~\bibnamefont {Schiffrin}},\ }\href@noop {} {\bibfield
  {journal} {\bibinfo  {journal} {Optica}\ }\textbf {\bibinfo {volume} {3}},\
  \bibinfo {pages} {1358} (\bibinfo {year} {2016})}\BibitemShut {NoStop}%
\bibitem [{\citenamefont {Wang}\ \emph {et~al.}(2012)\citenamefont {Wang},
  \citenamefont {Kalantar-Zadeh}, \citenamefont {Kis}, \citenamefont
  {Coleman},\ and\ \citenamefont
  {Strano}}]{Strano_et_al_nnano.2012.193_Transitional_Metal}%
  \BibitemOpen
  \bibfield  {author} {\bibinfo {author} {\bibfnamefont {Q.~H.}\ \bibnamefont
  {Wang}}, \bibinfo {author} {\bibfnamefont {K.}~\bibnamefont
  {Kalantar-Zadeh}}, \bibinfo {author} {\bibfnamefont {A.}~\bibnamefont {Kis}},
  \bibinfo {author} {\bibfnamefont {J.~N.}\ \bibnamefont {Coleman}}, \ and\
  \bibinfo {author} {\bibfnamefont {M.~S.}\ \bibnamefont {Strano}},\
  }\href@noop {} {\bibfield  {journal} {\bibinfo  {journal} {Nature
  Nanotechnology}\ }\textbf {\bibinfo {volume} {7}},\ \bibinfo {pages} {699}
  (\bibinfo {year} {2012})}\BibitemShut {NoStop}%
\bibitem [{\citenamefont {Britnell}\ \emph {et~al.}(2013)\citenamefont
  {Britnell}, \citenamefont {Ribeiro}, \citenamefont {Eckmann}, \citenamefont
  {Jalil}, \citenamefont {Belle}, \citenamefont {Mishchenko}, \citenamefont
  {Kim}, \citenamefont {Gorbachev}, \citenamefont {Georgiou}, \citenamefont
  {Morozov}, \citenamefont {Grigorenko}, \citenamefont {Geim}, \citenamefont
  {Casiraghi}, \citenamefont {Neto},\ and\ \citenamefont
  {Novoselov}}]{Novoselov_et_al_Science_2013_Light_Interaction_with_2D_Materials}%
  \BibitemOpen
  \bibfield  {author} {\bibinfo {author} {\bibfnamefont {L.}~\bibnamefont
  {Britnell}}, \bibinfo {author} {\bibfnamefont {R.~M.}\ \bibnamefont
  {Ribeiro}}, \bibinfo {author} {\bibfnamefont {A.}~\bibnamefont {Eckmann}},
  \bibinfo {author} {\bibfnamefont {R.}~\bibnamefont {Jalil}}, \bibinfo
  {author} {\bibfnamefont {B.~D.}\ \bibnamefont {Belle}}, \bibinfo {author}
  {\bibfnamefont {A.}~\bibnamefont {Mishchenko}}, \bibinfo {author}
  {\bibfnamefont {Y.~J.}\ \bibnamefont {Kim}}, \bibinfo {author} {\bibfnamefont
  {R.~V.}\ \bibnamefont {Gorbachev}}, \bibinfo {author} {\bibfnamefont
  {T.}~\bibnamefont {Georgiou}}, \bibinfo {author} {\bibfnamefont {S.~V.}\
  \bibnamefont {Morozov}}, \bibinfo {author} {\bibfnamefont {A.~N.}\
  \bibnamefont {Grigorenko}}, \bibinfo {author} {\bibfnamefont {A.~K.}\
  \bibnamefont {Geim}}, \bibinfo {author} {\bibfnamefont {C.}~\bibnamefont
  {Casiraghi}}, \bibinfo {author} {\bibfnamefont {A.~H.~C.}\ \bibnamefont
  {Neto}}, \ and\ \bibinfo {author} {\bibfnamefont {K.~S.}\ \bibnamefont
  {Novoselov}},\ }\href@noop {} {\bibfield  {journal} {\bibinfo  {journal}
  {Science}\ }\textbf {\bibinfo {volume} {340}},\ \bibinfo {pages} {1311}
  (\bibinfo {year} {2013})}\BibitemShut {NoStop}%
\bibitem [{\citenamefont {Withers}\ \emph {et~al.}(2015)\citenamefont
  {Withers}, \citenamefont {Pozo-Zamudio}, \citenamefont {Mishchenko},
  \citenamefont {Rooney}, \citenamefont {Gholinia}, \citenamefont {Watanabe},
  \citenamefont {Taniguchi}, \citenamefont {Haigh}, \citenamefont {Geim},
  \citenamefont {Tartakovskii},\ and\ \citenamefont
  {Novoselov}}]{Novoselov_et_al_Nat_Mat_2015_LED_of_2D_Heterostructures}%
  \BibitemOpen
  \bibfield  {author} {\bibinfo {author} {\bibfnamefont {F.}~\bibnamefont
  {Withers}}, \bibinfo {author} {\bibfnamefont {O.~D.}\ \bibnamefont
  {Pozo-Zamudio}}, \bibinfo {author} {\bibfnamefont {A.}~\bibnamefont
  {Mishchenko}}, \bibinfo {author} {\bibfnamefont {A.~P.}\ \bibnamefont
  {Rooney}}, \bibinfo {author} {\bibfnamefont {A.}~\bibnamefont {Gholinia}},
  \bibinfo {author} {\bibfnamefont {K.}~\bibnamefont {Watanabe}}, \bibinfo
  {author} {\bibfnamefont {T.}~\bibnamefont {Taniguchi}}, \bibinfo {author}
  {\bibfnamefont {S.~J.}\ \bibnamefont {Haigh}}, \bibinfo {author}
  {\bibfnamefont {A.~K.}\ \bibnamefont {Geim}}, \bibinfo {author}
  {\bibfnamefont {A.~I.}\ \bibnamefont {Tartakovskii}}, \ and\ \bibinfo
  {author} {\bibfnamefont {K.~S.}\ \bibnamefont {Novoselov}},\ }\href@noop {}
  {\bibfield  {journal} {\bibinfo  {journal} {Nat. Mater.}\ }\textbf {\bibinfo
  {volume} {14}},\ \bibinfo {pages} {301} (\bibinfo {year} {2015})}\BibitemShut
  {NoStop}%
\bibitem [{\citenamefont {Liu}\ \emph {et~al.}(2015)\citenamefont {Liu},
  \citenamefont {Xiao}, \citenamefont {Yao}, \citenamefont {Xu},\ and\
  \citenamefont
  {Yao}}]{Liu_et_al_Chemical_Society_Rev_2015_Electronic_structures}%
  \BibitemOpen
  \bibfield  {author} {\bibinfo {author} {\bibfnamefont {G.~B.}\ \bibnamefont
  {Liu}}, \bibinfo {author} {\bibfnamefont {D.}~\bibnamefont {Xiao}}, \bibinfo
  {author} {\bibfnamefont {Y.~G.}\ \bibnamefont {Yao}}, \bibinfo {author}
  {\bibfnamefont {X.~D.}\ \bibnamefont {Xu}}, \ and\ \bibinfo {author}
  {\bibfnamefont {W.}~\bibnamefont {Yao}},\ }\href@noop {} {\bibfield
  {journal} {\bibinfo  {journal} {Chem. Soc. Rev.}\ }\textbf {\bibinfo {volume}
  {44}},\ \bibinfo {pages} {2643} (\bibinfo {year} {2015})}\BibitemShut
  {NoStop}%
\bibitem [{\citenamefont {Novoselov}\ \emph {et~al.}(2016)\citenamefont
  {Novoselov}, \citenamefont {Mishchenko}, \citenamefont {Carvalho},\ and\
  \citenamefont
  {Neto}}]{Novoselov_et_al_Science_2016_2D_materials_and_van_der_Waals}%
  \BibitemOpen
  \bibfield  {author} {\bibinfo {author} {\bibfnamefont {K.~S.}\ \bibnamefont
  {Novoselov}}, \bibinfo {author} {\bibfnamefont {A.}~\bibnamefont
  {Mishchenko}}, \bibinfo {author} {\bibfnamefont {A.}~\bibnamefont
  {Carvalho}}, \ and\ \bibinfo {author} {\bibfnamefont {A.~H.~C.}\ \bibnamefont
  {Neto}},\ }\href@noop {} {\bibfield  {journal} {\bibinfo  {journal}
  {Science}\ }\textbf {\bibinfo {volume} {353}},\ \bibinfo {pages} {461}
  (\bibinfo {year} {2016})}\BibitemShut {NoStop}%
\bibitem [{\citenamefont {Ye}\ \emph {et~al.}(2016)\citenamefont {Ye},
  \citenamefont {Xiao}, \citenamefont {Wang}, \citenamefont {Ye}, \citenamefont
  {Zhu}, \citenamefont {Zhao}, \citenamefont {Wang}, \citenamefont {Zhao},
  \citenamefont {Yin},\ and\ \citenamefont
  {Zhang}}]{Ye_et_al_Nature_Nanotechnology_2016_Electrical_generation_and_control}%
  \BibitemOpen
  \bibfield  {author} {\bibinfo {author} {\bibfnamefont {Y.}~\bibnamefont
  {Ye}}, \bibinfo {author} {\bibfnamefont {J.}~\bibnamefont {Xiao}}, \bibinfo
  {author} {\bibfnamefont {H.~L.}\ \bibnamefont {Wang}}, \bibinfo {author}
  {\bibfnamefont {Z.~L.}\ \bibnamefont {Ye}}, \bibinfo {author} {\bibfnamefont
  {H.~Y.}\ \bibnamefont {Zhu}}, \bibinfo {author} {\bibfnamefont
  {M.}~\bibnamefont {Zhao}}, \bibinfo {author} {\bibfnamefont {Y.}~\bibnamefont
  {Wang}}, \bibinfo {author} {\bibfnamefont {J.~H.}\ \bibnamefont {Zhao}},
  \bibinfo {author} {\bibfnamefont {X.~B.}\ \bibnamefont {Yin}}, \ and\
  \bibinfo {author} {\bibfnamefont {X.}~\bibnamefont {Zhang}},\ }\href@noop {}
  {\bibfield  {journal} {\bibinfo  {journal} {Nature Nanotechnology}\ }\textbf
  {\bibinfo {volume} {11}},\ \bibinfo {pages} {598} (\bibinfo {year}
  {2016})}\BibitemShut {NoStop}%
\bibitem [{\citenamefont {Liu}\ \emph {et~al.}(2017)\citenamefont {Liu},
  \citenamefont {Li}, \citenamefont {You}, \citenamefont {Ghimire},
  \citenamefont {Heinz},\ and\ \citenamefont
  {Reis}}]{Reis_et_al_Nat_Phys_2017_HHG_from_2D_Crystals}%
  \BibitemOpen
  \bibfield  {author} {\bibinfo {author} {\bibfnamefont {H.~Z.}\ \bibnamefont
  {Liu}}, \bibinfo {author} {\bibfnamefont {Y.~L.}\ \bibnamefont {Li}},
  \bibinfo {author} {\bibfnamefont {Y.~S.}\ \bibnamefont {You}}, \bibinfo
  {author} {\bibfnamefont {S.}~\bibnamefont {Ghimire}}, \bibinfo {author}
  {\bibfnamefont {T.~F.}\ \bibnamefont {Heinz}}, \ and\ \bibinfo {author}
  {\bibfnamefont {D.~A.}\ \bibnamefont {Reis}},\ }\href@noop {} {\bibfield
  {journal} {\bibinfo  {journal} {Nat. Phys.}\ }\textbf {\bibinfo {volume}
  {13}},\ \bibinfo {pages} {262} (\bibinfo {year} {2017})}\BibitemShut
  {NoStop}%
\bibitem [{\citenamefont {Ashton}\ \emph {et~al.}(2017)\citenamefont {Ashton},
  \citenamefont {Paul}, \citenamefont {Sinnott},\ and\ \citenamefont
  {Hennig}}]{Ashton_et_al_PRL_2017_Topology_Scaling_Identification}%
  \BibitemOpen
  \bibfield  {author} {\bibinfo {author} {\bibfnamefont {M.}~\bibnamefont
  {Ashton}}, \bibinfo {author} {\bibfnamefont {J.}~\bibnamefont {Paul}},
  \bibinfo {author} {\bibfnamefont {S.~B.}\ \bibnamefont {Sinnott}}, \ and\
  \bibinfo {author} {\bibfnamefont {R.~G.}\ \bibnamefont {Hennig}},\
  }\href@noop {} {\bibfield  {journal} {\bibinfo  {journal} {Phys. Rev. Lett.}\
  }\textbf {\bibinfo {volume} {118}} (\bibinfo {year} {2017})}\BibitemShut
  {NoStop}%
\bibitem [{\citenamefont {Lemme}\ \emph {et~al.}(2014)\citenamefont {Lemme},
  \citenamefont {Li}, \citenamefont {Palacios},\ and\ \citenamefont
  {Schwierz}}]{lemme_li_palacios_schwierz_2014}%
  \BibitemOpen
  \bibfield  {author} {\bibinfo {author} {\bibfnamefont {M.~C.}\ \bibnamefont
  {Lemme}}, \bibinfo {author} {\bibfnamefont {L.-J.}\ \bibnamefont {Li}},
  \bibinfo {author} {\bibfnamefont {T.}~\bibnamefont {Palacios}}, \ and\
  \bibinfo {author} {\bibfnamefont {F.}~\bibnamefont {Schwierz}},\ }\href
  {\doibase 10.1557/mrs.2014.138} {\bibfield  {journal} {\bibinfo  {journal}
  {MRS Bulletin}\ }\textbf {\bibinfo {volume} {39}},\ \bibinfo {pages}
  {711–718} (\bibinfo {year} {2014})}\BibitemShut {NoStop}%
\bibitem [{\citenamefont {Novoselov}\ \emph {et~al.}(2012)\citenamefont
  {Novoselov}, \citenamefont {Falko}, \citenamefont {Colombo}, \citenamefont
  {Gellert}, \citenamefont {Schwab},\ and\ \citenamefont
  {Kim}}]{Novoselov_et_al_Nature_2012_Graphene_Review}%
  \BibitemOpen
  \bibfield  {author} {\bibinfo {author} {\bibfnamefont {K.~S.}\ \bibnamefont
  {Novoselov}}, \bibinfo {author} {\bibfnamefont {V.~I.}\ \bibnamefont
  {Falko}}, \bibinfo {author} {\bibfnamefont {L.}~\bibnamefont {Colombo}},
  \bibinfo {author} {\bibfnamefont {P.~R.}\ \bibnamefont {Gellert}}, \bibinfo
  {author} {\bibfnamefont {M.~G.}\ \bibnamefont {Schwab}}, \ and\ \bibinfo
  {author} {\bibfnamefont {K.}~\bibnamefont {Kim}},\ }\href@noop {} {\bibfield
  {journal} {\bibinfo  {journal} {Nature}\ }\textbf {\bibinfo {volume} {490}},\
  \bibinfo {pages} {192} (\bibinfo {year} {2012})}\BibitemShut {NoStop}%
\bibitem [{\citenamefont {Jariwala}\ \emph {et~al.}(2014)\citenamefont
  {Jariwala}, \citenamefont {Sangwan}, \citenamefont {Lauhon}, \citenamefont
  {Marks},\ and\ \citenamefont
  {Hersam}}]{Jariwala_et_al_Asc_Nano_2014_Transition_Metal}%
  \BibitemOpen
  \bibfield  {author} {\bibinfo {author} {\bibfnamefont {D.}~\bibnamefont
  {Jariwala}}, \bibinfo {author} {\bibfnamefont {V.~K.}\ \bibnamefont
  {Sangwan}}, \bibinfo {author} {\bibfnamefont {L.~J.}\ \bibnamefont {Lauhon}},
  \bibinfo {author} {\bibfnamefont {T.~J.}\ \bibnamefont {Marks}}, \ and\
  \bibinfo {author} {\bibfnamefont {M.~C.}\ \bibnamefont {Hersam}},\
  }\href@noop {} {\bibfield  {journal} {\bibinfo  {journal} {Acs Nano}\
  }\textbf {\bibinfo {volume} {8}},\ \bibinfo {pages} {1102} (\bibinfo {year}
  {2014})}\BibitemShut {NoStop}%
\bibitem [{\citenamefont {Geim}\ and\ \citenamefont
  {Novoselov}(2007)}]{Geim_et_al_Nat_Mater_2007_The_rise_of_graphene}%
  \BibitemOpen
  \bibfield  {author} {\bibinfo {author} {\bibfnamefont {A.~K.}\ \bibnamefont
  {Geim}}\ and\ \bibinfo {author} {\bibfnamefont {K.~S.}\ \bibnamefont
  {Novoselov}},\ }\href@noop {} {\bibfield  {journal} {\bibinfo  {journal} {Nat
  Mater}\ }\textbf {\bibinfo {volume} {6}},\ \bibinfo {pages} {183} (\bibinfo
  {year} {2007})}\BibitemShut {NoStop}%
\bibitem [{\citenamefont
  {Schwierz}(2010)}]{Schwierz_Nature_Nanotechnology_2010_Graphene_transistors}%
  \BibitemOpen
  \bibfield  {author} {\bibinfo {author} {\bibfnamefont {F.}~\bibnamefont
  {Schwierz}},\ }\href@noop {} {\bibfield  {journal} {\bibinfo  {journal}
  {Nature Nanotechnology}\ }\textbf {\bibinfo {volume} {5}},\ \bibinfo {pages}
  {487} (\bibinfo {year} {2010})}\BibitemShut {NoStop}%
\bibitem [{\citenamefont {Liu}\ \emph {et~al.}(2014)\citenamefont {Liu},
  \citenamefont {Shan}, \citenamefont {Yao}, \citenamefont {Yao},\ and\
  \citenamefont {Xiao}}]{Liu_et_al_PRB_2014_Three_Band_Model}%
  \BibitemOpen
  \bibfield  {author} {\bibinfo {author} {\bibfnamefont {G.~B.}\ \bibnamefont
  {Liu}}, \bibinfo {author} {\bibfnamefont {W.~Y.}\ \bibnamefont {Shan}},
  \bibinfo {author} {\bibfnamefont {Y.~G.}\ \bibnamefont {Yao}}, \bibinfo
  {author} {\bibfnamefont {W.}~\bibnamefont {Yao}}, \ and\ \bibinfo {author}
  {\bibfnamefont {D.}~\bibnamefont {Xiao}},\ }\href@noop {} {\bibfield
  {journal} {\bibinfo  {journal} {Phys. Rev. B}\ }\textbf {\bibinfo {volume}
  {89}} (\bibinfo {year} {2014})}\BibitemShut {NoStop}%
\bibitem [{\citenamefont {Xiao}\ \emph {et~al.}(2012)\citenamefont {Xiao},
  \citenamefont {Liu}, \citenamefont {Feng}, \citenamefont {Xu},\ and\
  \citenamefont {Yao}}]{Xiao_et_al_PRL_2012_Coupled_Spin_and_Valley_Physics}%
  \BibitemOpen
  \bibfield  {author} {\bibinfo {author} {\bibfnamefont {D.}~\bibnamefont
  {Xiao}}, \bibinfo {author} {\bibfnamefont {G.~B.}\ \bibnamefont {Liu}},
  \bibinfo {author} {\bibfnamefont {W.~X.}\ \bibnamefont {Feng}}, \bibinfo
  {author} {\bibfnamefont {X.~D.}\ \bibnamefont {Xu}}, \ and\ \bibinfo {author}
  {\bibfnamefont {W.}~\bibnamefont {Yao}},\ }\href@noop {} {\bibfield
  {journal} {\bibinfo  {journal} {Phys. Rev. Lett.}\ }\textbf {\bibinfo
  {volume} {108}} (\bibinfo {year} {2012})}\BibitemShut {NoStop}%
\bibitem [{\citenamefont
  {Jiang}(2015)}]{Jiang_Frontiers_of_Physics_2015_Graphene_versus_MoS2}%
  \BibitemOpen
  \bibfield  {author} {\bibinfo {author} {\bibfnamefont {J.~W.}\ \bibnamefont
  {Jiang}},\ }\href@noop {} {\bibfield  {journal} {\bibinfo  {journal}
  {Frontiers of Physics}\ }\textbf {\bibinfo {volume} {10}},\ \bibinfo {pages}
  {287} (\bibinfo {year} {2015})}\BibitemShut {NoStop}%
\bibitem [{\citenamefont {Sie}\ \emph {et~al.}(2015)\citenamefont {Sie},
  \citenamefont {McIver}, \citenamefont {Lee}, \citenamefont {Fu},
  \citenamefont {Kong},\ and\ \citenamefont
  {Gedik}}]{Sie_et_al_Nature_Materials_2015_Valley-selective_optical}%
  \BibitemOpen
  \bibfield  {author} {\bibinfo {author} {\bibfnamefont {E.~J.}\ \bibnamefont
  {Sie}}, \bibinfo {author} {\bibfnamefont {J.}~\bibnamefont {McIver}},
  \bibinfo {author} {\bibfnamefont {Y.~H.}\ \bibnamefont {Lee}}, \bibinfo
  {author} {\bibfnamefont {L.}~\bibnamefont {Fu}}, \bibinfo {author}
  {\bibfnamefont {J.}~\bibnamefont {Kong}}, \ and\ \bibinfo {author}
  {\bibfnamefont {N.}~\bibnamefont {Gedik}},\ }\href@noop {} {\bibfield
  {journal} {\bibinfo  {journal} {Nature Materials}\ }\textbf {\bibinfo
  {volume} {14}},\ \bibinfo {pages} {290} (\bibinfo {year} {2015})}\BibitemShut
  {NoStop}%
\bibitem [{\citenamefont {Zeng}\ \emph {et~al.}(2012)\citenamefont {Zeng},
  \citenamefont {Dai}, \citenamefont {Yao}, \citenamefont {Xiao},\ and\
  \citenamefont
  {Cui}}]{Zeng_et_al_Nature_Nanotechnology_2012_Valley_polarization_in_MoS2}%
  \BibitemOpen
  \bibfield  {author} {\bibinfo {author} {\bibfnamefont {H.~L.}\ \bibnamefont
  {Zeng}}, \bibinfo {author} {\bibfnamefont {J.~F.}\ \bibnamefont {Dai}},
  \bibinfo {author} {\bibfnamefont {W.}~\bibnamefont {Yao}}, \bibinfo {author}
  {\bibfnamefont {D.}~\bibnamefont {Xiao}}, \ and\ \bibinfo {author}
  {\bibfnamefont {X.~D.}\ \bibnamefont {Cui}},\ }\href@noop {} {\bibfield
  {journal} {\bibinfo  {journal} {Nature Nanotechnology}\ }\textbf {\bibinfo
  {volume} {7}},\ \bibinfo {pages} {490} (\bibinfo {year} {2012})}\BibitemShut
  {NoStop}%
\bibitem [{\citenamefont {Mak}\ \emph {et~al.}(2012)\citenamefont {Mak},
  \citenamefont {He}, \citenamefont {Shan},\ and\ \citenamefont
  {Heinz}}]{Heinz_et_al_10.1038_Nnano.2012.96_Valley_Polarization_in_TMDC_by_Optical_Helicity}%
  \BibitemOpen
  \bibfield  {author} {\bibinfo {author} {\bibfnamefont {K.~F.}\ \bibnamefont
  {Mak}}, \bibinfo {author} {\bibfnamefont {K.~L.}\ \bibnamefont {He}},
  \bibinfo {author} {\bibfnamefont {J.}~\bibnamefont {Shan}}, \ and\ \bibinfo
  {author} {\bibfnamefont {T.~F.}\ \bibnamefont {Heinz}},\ }\href@noop {}
  {\bibfield  {journal} {\bibinfo  {journal} {Nature Nanotechnology}\ }\textbf
  {\bibinfo {volume} {7}},\ \bibinfo {pages} {494} (\bibinfo {year}
  {2012})}\BibitemShut {NoStop}%
\bibitem [{\citenamefont {Cao}\ \emph {et~al.}(2012)\citenamefont {Cao},
  \citenamefont {Wang}, \citenamefont {Han}, \citenamefont {Ye}, \citenamefont
  {Zhu}, \citenamefont {Shi}, \citenamefont {Niu}, \citenamefont {Tan},
  \citenamefont {Wang}, \citenamefont {Liu},\ and\ \citenamefont
  {Feng}}]{Feng_et_al_ncomms1882_2012_Valley_Selective_CD}%
  \BibitemOpen
  \bibfield  {author} {\bibinfo {author} {\bibfnamefont {T.}~\bibnamefont
  {Cao}}, \bibinfo {author} {\bibfnamefont {G.}~\bibnamefont {Wang}}, \bibinfo
  {author} {\bibfnamefont {W.~P.}\ \bibnamefont {Han}}, \bibinfo {author}
  {\bibfnamefont {H.~Q.}\ \bibnamefont {Ye}}, \bibinfo {author} {\bibfnamefont
  {C.~R.}\ \bibnamefont {Zhu}}, \bibinfo {author} {\bibfnamefont {J.~R.}\
  \bibnamefont {Shi}}, \bibinfo {author} {\bibfnamefont {Q.}~\bibnamefont
  {Niu}}, \bibinfo {author} {\bibfnamefont {P.~H.}\ \bibnamefont {Tan}},
  \bibinfo {author} {\bibfnamefont {E.}~\bibnamefont {Wang}}, \bibinfo {author}
  {\bibfnamefont {B.~L.}\ \bibnamefont {Liu}}, \ and\ \bibinfo {author}
  {\bibfnamefont {J.}~\bibnamefont {Feng}},\ }\href@noop {} {\bibfield
  {journal} {\bibinfo  {journal} {Nat. Commun.}\ }\textbf {\bibinfo {volume}
  {3}},\ \bibinfo {pages} {887} (\bibinfo {year} {2012})}\BibitemShut {NoStop}%
\bibitem [{\citenamefont {Jones}\ \emph {et~al.}(2013)\citenamefont {Jones},
  \citenamefont {Yu}, \citenamefont {Ghimire}, \citenamefont {Wu},
  \citenamefont {Aivazian}, \citenamefont {Ross}, \citenamefont {Zhao},
  \citenamefont {Yan}, \citenamefont {Mandrus}, \citenamefont {Xiao},
  \citenamefont {Yao},\ and\ \citenamefont
  {Xu}}]{Jones_et_al_Nature_Nanotechnology_2013_Optical_generation_of_excitonic_valley}%
  \BibitemOpen
  \bibfield  {author} {\bibinfo {author} {\bibfnamefont {A.~M.}\ \bibnamefont
  {Jones}}, \bibinfo {author} {\bibfnamefont {H.~Y.}\ \bibnamefont {Yu}},
  \bibinfo {author} {\bibfnamefont {N.~J.}\ \bibnamefont {Ghimire}}, \bibinfo
  {author} {\bibfnamefont {S.~F.}\ \bibnamefont {Wu}}, \bibinfo {author}
  {\bibfnamefont {G.}~\bibnamefont {Aivazian}}, \bibinfo {author}
  {\bibfnamefont {J.~S.}\ \bibnamefont {Ross}}, \bibinfo {author}
  {\bibfnamefont {B.}~\bibnamefont {Zhao}}, \bibinfo {author} {\bibfnamefont
  {J.~Q.}\ \bibnamefont {Yan}}, \bibinfo {author} {\bibfnamefont {D.~G.}\
  \bibnamefont {Mandrus}}, \bibinfo {author} {\bibfnamefont {D.}~\bibnamefont
  {Xiao}}, \bibinfo {author} {\bibfnamefont {W.}~\bibnamefont {Yao}}, \ and\
  \bibinfo {author} {\bibfnamefont {X.~D.}\ \bibnamefont {Xu}},\ }\href@noop {}
  {\bibfield  {journal} {\bibinfo  {journal} {Nature Nanotechnology}\ }\textbf
  {\bibinfo {volume} {8}},\ \bibinfo {pages} {634} (\bibinfo {year}
  {2013})}\BibitemShut {NoStop}%
\bibitem [{\citenamefont {Eginligil}\ \emph {et~al.}(2015)\citenamefont
  {Eginligil}, \citenamefont {Cao}, \citenamefont {Wang}, \citenamefont {Shen},
  \citenamefont {Cong}, \citenamefont {Shang}, \citenamefont {Soci},\ and\
  \citenamefont
  {Yu}}]{Eginligil_et_al_Nature_Communications_2015_Dichroic_spin_valley_photocurrent}%
  \BibitemOpen
  \bibfield  {author} {\bibinfo {author} {\bibfnamefont {M.}~\bibnamefont
  {Eginligil}}, \bibinfo {author} {\bibfnamefont {B.~C.}\ \bibnamefont {Cao}},
  \bibinfo {author} {\bibfnamefont {Z.~L.}\ \bibnamefont {Wang}}, \bibinfo
  {author} {\bibfnamefont {X.~N.}\ \bibnamefont {Shen}}, \bibinfo {author}
  {\bibfnamefont {C.~X.}\ \bibnamefont {Cong}}, \bibinfo {author}
  {\bibfnamefont {J.~Z.}\ \bibnamefont {Shang}}, \bibinfo {author}
  {\bibfnamefont {C.}~\bibnamefont {Soci}}, \ and\ \bibinfo {author}
  {\bibfnamefont {T.}~\bibnamefont {Yu}},\ }\href@noop {} {\bibfield  {journal}
  {\bibinfo  {journal} {Nature Communications}\ }\textbf {\bibinfo {volume}
  {6}} (\bibinfo {year} {2015})}\BibitemShut {NoStop}%
\bibitem [{\citenamefont {Garg}\ \emph {et~al.}(2016)\citenamefont {Garg},
  \citenamefont {Zhan}, \citenamefont {Luu}, \citenamefont {Lakhotia},
  \citenamefont {Klostermann}, \citenamefont {Guggenmos},\ and\ \citenamefont
  {Goulielmakis}}]{Garg-Nature_2016-Multi-petahertz_electronic_metrology}%
  \BibitemOpen
  \bibfield  {author} {\bibinfo {author} {\bibfnamefont {M.}~\bibnamefont
  {Garg}}, \bibinfo {author} {\bibfnamefont {M.}~\bibnamefont {Zhan}}, \bibinfo
  {author} {\bibfnamefont {T.~T.}\ \bibnamefont {Luu}}, \bibinfo {author}
  {\bibfnamefont {H.}~\bibnamefont {Lakhotia}}, \bibinfo {author}
  {\bibfnamefont {T.}~\bibnamefont {Klostermann}}, \bibinfo {author}
  {\bibfnamefont {A.}~\bibnamefont {Guggenmos}}, \ and\ \bibinfo {author}
  {\bibfnamefont {E.}~\bibnamefont {Goulielmakis}},\ }\href@noop {} {\bibfield
  {journal} {\bibinfo  {journal} {Nature}\ }\textbf {\bibinfo {volume} {538}},\
  \bibinfo {pages} {359} (\bibinfo {year} {2016})}\BibitemShut {NoStop}%
\bibitem [{\citenamefont {Krogen}\ \emph {et~al.}(2017)\citenamefont {Krogen},
  \citenamefont {Suchowski}, \citenamefont {Liang}, \citenamefont {Flemens},
  \citenamefont {Hong}, \citenamefont {Kaertner},\ and\ \citenamefont
  {Moses}}]{Krogen-2017-Generation_and_multi-octave_shaping_in_mid-IR}%
  \BibitemOpen
  \bibfield  {author} {\bibinfo {author} {\bibfnamefont {P.}~\bibnamefont
  {Krogen}}, \bibinfo {author} {\bibfnamefont {H.}~\bibnamefont {Suchowski}},
  \bibinfo {author} {\bibfnamefont {H.}~\bibnamefont {Liang}}, \bibinfo
  {author} {\bibfnamefont {N.}~\bibnamefont {Flemens}}, \bibinfo {author}
  {\bibfnamefont {K.-H.}\ \bibnamefont {Hong}}, \bibinfo {author}
  {\bibfnamefont {F.~X.}\ \bibnamefont {Kaertner}}, \ and\ \bibinfo {author}
  {\bibfnamefont {J.}~\bibnamefont {Moses}},\ }\href@noop {} {\bibfield
  {journal} {\bibinfo  {journal} {Nat. Phot.}\ }\textbf {\bibinfo {volume}
  {11}},\ \bibinfo {pages} {222} (\bibinfo {year} {2017})}\BibitemShut
  {NoStop}%
\bibitem [{\citenamefont {Liang}\ \emph {et~al.}(2017)\citenamefont {Liang},
  \citenamefont {Krogen}, \citenamefont {Wang}, \citenamefont {Park},
  \citenamefont {Kroh}, \citenamefont {Zawilski}, \citenamefont {Schunemann},
  \citenamefont {Moses}, \citenamefont {DiMauro}, \citenamefont {Kaertner},\
  and\ \citenamefont
  {Hong}}]{Liang-Nat_Commun_2017-High-energy_mid-infrared_sub-cycle}%
  \BibitemOpen
  \bibfield  {author} {\bibinfo {author} {\bibfnamefont {H.}~\bibnamefont
  {Liang}}, \bibinfo {author} {\bibfnamefont {P.}~\bibnamefont {Krogen}},
  \bibinfo {author} {\bibfnamefont {Z.}~\bibnamefont {Wang}}, \bibinfo {author}
  {\bibfnamefont {H.}~\bibnamefont {Park}}, \bibinfo {author} {\bibfnamefont
  {T.}~\bibnamefont {Kroh}}, \bibinfo {author} {\bibfnamefont {K.}~\bibnamefont
  {Zawilski}}, \bibinfo {author} {\bibfnamefont {P.}~\bibnamefont
  {Schunemann}}, \bibinfo {author} {\bibfnamefont {J.}~\bibnamefont {Moses}},
  \bibinfo {author} {\bibfnamefont {L.~F.}\ \bibnamefont {DiMauro}}, \bibinfo
  {author} {\bibfnamefont {F.~X.}\ \bibnamefont {Kaertner}}, \ and\ \bibinfo
  {author} {\bibfnamefont {K.-H.}\ \bibnamefont {Hong}},\ }\href@noop {}
  {\bibfield  {journal} {\bibinfo  {journal} {Nat. Commun.}\ }\textbf {\bibinfo
  {volume} {8}},\ \bibinfo {pages} {141} (\bibinfo {year} {2017})}\BibitemShut
  {NoStop}%
\bibitem [{\citenamefont {Elu}\ \emph {et~al.}(2017)\citenamefont {Elu},
  \citenamefont {Baudisch}, \citenamefont {Pires}, \citenamefont {Tani},
  \citenamefont {Frosz}, \citenamefont {Koettig}, \citenamefont {Ermolov},
  \citenamefont {Russell},\ and\ \citenamefont
  {Biegert}}]{Russel_et_al_optica-4-9-1024_2017_Single_Cycle_MidIR_Pulses_from_OPCPA}%
  \BibitemOpen
  \bibfield  {author} {\bibinfo {author} {\bibfnamefont {U.}~\bibnamefont
  {Elu}}, \bibinfo {author} {\bibfnamefont {M.}~\bibnamefont {Baudisch}},
  \bibinfo {author} {\bibfnamefont {H.}~\bibnamefont {Pires}}, \bibinfo
  {author} {\bibfnamefont {F.}~\bibnamefont {Tani}}, \bibinfo {author}
  {\bibfnamefont {M.~H.}\ \bibnamefont {Frosz}}, \bibinfo {author}
  {\bibfnamefont {F.}~\bibnamefont {Koettig}}, \bibinfo {author} {\bibfnamefont
  {A.}~\bibnamefont {Ermolov}}, \bibinfo {author} {\bibfnamefont {P.~S.}\
  \bibnamefont {Russell}}, \ and\ \bibinfo {author} {\bibfnamefont
  {J.}~\bibnamefont {Biegert}},\ }\href@noop {} {\bibfield  {journal} {\bibinfo
   {journal} {Optica}\ }\textbf {\bibinfo {volume} {4}},\ \bibinfo {pages}
  {1024} (\bibinfo {year} {2017})}\BibitemShut {NoStop}%
\bibitem [{\citenamefont {Vicario}\ \emph {et~al.}(2017)\citenamefont
  {Vicario}, \citenamefont {Shalaby},\ and\ \citenamefont
  {Hauri}}]{Hauri_PhysRevLett.118_2017_Subcycle_THz_Semiconductors}%
  \BibitemOpen
  \bibfield  {author} {\bibinfo {author} {\bibfnamefont {C.}~\bibnamefont
  {Vicario}}, \bibinfo {author} {\bibfnamefont {M.}~\bibnamefont {Shalaby}}, \
  and\ \bibinfo {author} {\bibfnamefont {C.~P.}\ \bibnamefont {Hauri}},\
  }\href@noop {} {\bibfield  {journal} {\bibinfo  {journal} {Phys. Rev. Lett.}\
  }\textbf {\bibinfo {volume} {118}},\ \bibinfo {pages} {083901} (\bibinfo
  {year} {2017})}\BibitemShut {NoStop}%
\bibitem [{\citenamefont {Dhillon}\ \emph {et~al.}(2017)\citenamefont
  {Dhillon}, \citenamefont {Vitiello}, \citenamefont {Linfield}, \citenamefont
  {Davies}, \citenamefont {Hoffmann}, \citenamefont {Booske}, \citenamefont
  {Paoloni}, \citenamefont {Gensch}, \citenamefont {Weightman}, \citenamefont
  {Williams}, \citenamefont {Castro-Camus}, \citenamefont {Cumming},
  \citenamefont {Simoens}, \citenamefont {Escorcia-Carranza}, \citenamefont
  {Grant}, \citenamefont {Lucyszyn}, \citenamefont {Kuwata-Gonokami},
  \citenamefont {Konishi}, \citenamefont {Koch}, \citenamefont {Schmuttenmaer},
  \citenamefont {Cocker}, \citenamefont {Huber}, \citenamefont {Markelz},
  \citenamefont {Taylor}, \citenamefont {Wallace}, \citenamefont {Zeitler},
  \citenamefont {Sibik}, \citenamefont {Korter}, \citenamefont {Ellison},
  \citenamefont {Rea}, \citenamefont {Goldsmith}, \citenamefont {Cooper},
  \citenamefont {Appleby}, \citenamefont {Pardo}, \citenamefont {Huggard},
  \citenamefont {Krozer}, \citenamefont {Shams}, \citenamefont {Fice},
  \citenamefont {Renaud}, \citenamefont {Seeds}, \citenamefont {Stohr},
  \citenamefont {Naftaly}, \citenamefont {Ridler}, \citenamefont {Clarke},
  \citenamefont {Cunningham},\ and\ \citenamefont
  {Johnston}}]{Dhillon_2017_terahertz_science_roadmap}%
  \BibitemOpen
  \bibfield  {author} {\bibinfo {author} {\bibfnamefont {S.~S.}\ \bibnamefont
  {Dhillon}}, \bibinfo {author} {\bibfnamefont {M.~S.}\ \bibnamefont
  {Vitiello}}, \bibinfo {author} {\bibfnamefont {E.~H.}\ \bibnamefont
  {Linfield}}, \bibinfo {author} {\bibfnamefont {A.~G.}\ \bibnamefont
  {Davies}}, \bibinfo {author} {\bibfnamefont {M.~C.}\ \bibnamefont
  {Hoffmann}}, \bibinfo {author} {\bibfnamefont {J.}~\bibnamefont {Booske}},
  \bibinfo {author} {\bibfnamefont {C.}~\bibnamefont {Paoloni}}, \bibinfo
  {author} {\bibfnamefont {M.}~\bibnamefont {Gensch}}, \bibinfo {author}
  {\bibfnamefont {P.}~\bibnamefont {Weightman}}, \bibinfo {author}
  {\bibfnamefont {G.~P.}\ \bibnamefont {Williams}}, \bibinfo {author}
  {\bibfnamefont {E.}~\bibnamefont {Castro-Camus}}, \bibinfo {author}
  {\bibfnamefont {D.~R.~S.}\ \bibnamefont {Cumming}}, \bibinfo {author}
  {\bibfnamefont {F.}~\bibnamefont {Simoens}}, \bibinfo {author} {\bibfnamefont
  {I.}~\bibnamefont {Escorcia-Carranza}}, \bibinfo {author} {\bibfnamefont
  {J.}~\bibnamefont {Grant}}, \bibinfo {author} {\bibfnamefont
  {S.}~\bibnamefont {Lucyszyn}}, \bibinfo {author} {\bibfnamefont
  {M.}~\bibnamefont {Kuwata-Gonokami}}, \bibinfo {author} {\bibfnamefont
  {K.}~\bibnamefont {Konishi}}, \bibinfo {author} {\bibfnamefont
  {M.}~\bibnamefont {Koch}}, \bibinfo {author} {\bibfnamefont {C.~A.}\
  \bibnamefont {Schmuttenmaer}}, \bibinfo {author} {\bibfnamefont {T.~L.}\
  \bibnamefont {Cocker}}, \bibinfo {author} {\bibfnamefont {R.}~\bibnamefont
  {Huber}}, \bibinfo {author} {\bibfnamefont {A.~G.}\ \bibnamefont {Markelz}},
  \bibinfo {author} {\bibfnamefont {Z.~D.}\ \bibnamefont {Taylor}}, \bibinfo
  {author} {\bibfnamefont {V.~P.}\ \bibnamefont {Wallace}}, \bibinfo {author}
  {\bibfnamefont {J.~A.}\ \bibnamefont {Zeitler}}, \bibinfo {author}
  {\bibfnamefont {J.}~\bibnamefont {Sibik}}, \bibinfo {author} {\bibfnamefont
  {T.~M.}\ \bibnamefont {Korter}}, \bibinfo {author} {\bibfnamefont
  {B.}~\bibnamefont {Ellison}}, \bibinfo {author} {\bibfnamefont
  {S.}~\bibnamefont {Rea}}, \bibinfo {author} {\bibfnamefont {P.}~\bibnamefont
  {Goldsmith}}, \bibinfo {author} {\bibfnamefont {K.~B.}\ \bibnamefont
  {Cooper}}, \bibinfo {author} {\bibfnamefont {R.}~\bibnamefont {Appleby}},
  \bibinfo {author} {\bibfnamefont {D.}~\bibnamefont {Pardo}}, \bibinfo
  {author} {\bibfnamefont {P.~G.}\ \bibnamefont {Huggard}}, \bibinfo {author}
  {\bibfnamefont {V.}~\bibnamefont {Krozer}}, \bibinfo {author} {\bibfnamefont
  {H.}~\bibnamefont {Shams}}, \bibinfo {author} {\bibfnamefont
  {M.}~\bibnamefont {Fice}}, \bibinfo {author} {\bibfnamefont {C.}~\bibnamefont
  {Renaud}}, \bibinfo {author} {\bibfnamefont {A.}~\bibnamefont {Seeds}},
  \bibinfo {author} {\bibfnamefont {A.}~\bibnamefont {Stohr}}, \bibinfo
  {author} {\bibfnamefont {M.}~\bibnamefont {Naftaly}}, \bibinfo {author}
  {\bibfnamefont {N.}~\bibnamefont {Ridler}}, \bibinfo {author} {\bibfnamefont
  {R.}~\bibnamefont {Clarke}}, \bibinfo {author} {\bibfnamefont {J.~E.}\
  \bibnamefont {Cunningham}}, \ and\ \bibinfo {author} {\bibfnamefont {M.~B.}\
  \bibnamefont {Johnston}},\ }\href@noop {} {\bibfield  {journal} {\bibinfo
  {journal} {J. Phys. D Appl. Phys.}\ }\textbf {\bibinfo {volume} {50}},\
  \bibinfo {pages} {043001} (\bibinfo {year} {2017})}\BibitemShut {NoStop}%
\bibitem [{\citenamefont {Langer}\ \emph {et~al.}(2017)\citenamefont {Langer},
  \citenamefont {Hohenleutner}, \citenamefont {Huttner}, \citenamefont {Koch},
  \citenamefont {Kira},\ and\ \citenamefont
  {Huber}}]{Langer-2017-Symmetry-controlled_THz}%
  \BibitemOpen
  \bibfield  {author} {\bibinfo {author} {\bibfnamefont {F.}~\bibnamefont
  {Langer}}, \bibinfo {author} {\bibfnamefont {M.}~\bibnamefont
  {Hohenleutner}}, \bibinfo {author} {\bibfnamefont {U.}~\bibnamefont
  {Huttner}}, \bibinfo {author} {\bibfnamefont {S.~W.}\ \bibnamefont {Koch}},
  \bibinfo {author} {\bibfnamefont {M.}~\bibnamefont {Kira}}, \ and\ \bibinfo
  {author} {\bibfnamefont {R.}~\bibnamefont {Huber}},\ }\href@noop {}
  {\bibfield  {journal} {\bibinfo  {journal} {Nat. Phot.}\ }\textbf {\bibinfo
  {volume} {11}},\ \bibinfo {pages} {227} (\bibinfo {year} {2017})}\BibitemShut
  {NoStop}%
\bibitem [{\citenamefont {Li}\ \emph {et~al.}(2016)\citenamefont {Li},
  \citenamefont {Ren}, \citenamefont {Yin}, \citenamefont {Cheng},
  \citenamefont {Cunningham}, \citenamefont {Wu},\ and\ \citenamefont
  {Chang}}]{Li-2016-Polarization_gating_of_HHG}%
  \BibitemOpen
  \bibfield  {author} {\bibinfo {author} {\bibfnamefont {J.}~\bibnamefont
  {Li}}, \bibinfo {author} {\bibfnamefont {X.}~\bibnamefont {Ren}}, \bibinfo
  {author} {\bibfnamefont {Y.}~\bibnamefont {Yin}}, \bibinfo {author}
  {\bibfnamefont {Y.}~\bibnamefont {Cheng}}, \bibinfo {author} {\bibfnamefont
  {E.}~\bibnamefont {Cunningham}}, \bibinfo {author} {\bibfnamefont
  {Y.}~\bibnamefont {Wu}}, \ and\ \bibinfo {author} {\bibfnamefont
  {Z.}~\bibnamefont {Chang}},\ }\href@noop {} {\bibfield  {journal} {\bibinfo
  {journal} {Appl. Phys. Lett.}\ }\textbf {\bibinfo {volume} {108}},\ \bibinfo
  {pages} {231102} (\bibinfo {year} {2016})}\BibitemShut {NoStop}%
\bibitem [{\citenamefont {Sim}\ \emph {et~al.}(2013)\citenamefont {Sim},
  \citenamefont {Park}, \citenamefont {Song}, \citenamefont {In}, \citenamefont
  {Lee}, \citenamefont {Kim},\ and\ \citenamefont
  {Choi}}]{Sim_et_al_PRB_2013_Exciton_dynamics_in_atomically_thin_MoS2}%
  \BibitemOpen
  \bibfield  {author} {\bibinfo {author} {\bibfnamefont {S.}~\bibnamefont
  {Sim}}, \bibinfo {author} {\bibfnamefont {J.}~\bibnamefont {Park}}, \bibinfo
  {author} {\bibfnamefont {J.~G.}\ \bibnamefont {Song}}, \bibinfo {author}
  {\bibfnamefont {C.}~\bibnamefont {In}}, \bibinfo {author} {\bibfnamefont
  {Y.~S.}\ \bibnamefont {Lee}}, \bibinfo {author} {\bibfnamefont
  {H.}~\bibnamefont {Kim}}, \ and\ \bibinfo {author} {\bibfnamefont
  {H.}~\bibnamefont {Choi}},\ }\href@noop {} {\bibfield  {journal} {\bibinfo
  {journal} {Phys. Rev. B}\ }\textbf {\bibinfo {volume} {88}} (\bibinfo {year}
  {2013})}\BibitemShut {NoStop}%
\bibitem [{\citenamefont {Moody}\ \emph {et~al.}(2016)\citenamefont {Moody},
  \citenamefont {Schaibley},\ and\ \citenamefont
  {Xu}}]{Moody_et_al_Journal_of_the_Optical_Society_2016_Exciton_dynamics_in_monolayer}%
  \BibitemOpen
  \bibfield  {author} {\bibinfo {author} {\bibfnamefont {G.}~\bibnamefont
  {Moody}}, \bibinfo {author} {\bibfnamefont {J.}~\bibnamefont {Schaibley}}, \
  and\ \bibinfo {author} {\bibfnamefont {X.~D.}\ \bibnamefont {Xu}},\
  }\href@noop {} {\bibfield  {journal} {\bibinfo  {journal} {Journal of the
  Optical Society of America B-Optical Physics}\ }\textbf {\bibinfo {volume}
  {33}},\ \bibinfo {pages} {C39} (\bibinfo {year} {2016})}\BibitemShut
  {NoStop}%
\bibitem [{\citenamefont {Nie}\ \emph {et~al.}(2014)\citenamefont {Nie},
  \citenamefont {Long}, \citenamefont {Sun}, \citenamefont {Huang},
  \citenamefont {Zhang}, \citenamefont {Xiong}, \citenamefont {Hewak},
  \citenamefont {Shen}, \citenamefont {Prezhdo},\ and\ \citenamefont
  {Loh}}]{Nie_et_al_Acs_Nano_2014_Ultrafast_Carrier_Thermalization}%
  \BibitemOpen
  \bibfield  {author} {\bibinfo {author} {\bibfnamefont {Z.~G.}\ \bibnamefont
  {Nie}}, \bibinfo {author} {\bibfnamefont {R.}~\bibnamefont {Long}}, \bibinfo
  {author} {\bibfnamefont {L.~F.}\ \bibnamefont {Sun}}, \bibinfo {author}
  {\bibfnamefont {C.~C.}\ \bibnamefont {Huang}}, \bibinfo {author}
  {\bibfnamefont {J.}~\bibnamefont {Zhang}}, \bibinfo {author} {\bibfnamefont
  {Q.~H.}\ \bibnamefont {Xiong}}, \bibinfo {author} {\bibfnamefont {D.~W.}\
  \bibnamefont {Hewak}}, \bibinfo {author} {\bibfnamefont {Z.~X.}\ \bibnamefont
  {Shen}}, \bibinfo {author} {\bibfnamefont {O.~V.}\ \bibnamefont {Prezhdo}}, \
  and\ \bibinfo {author} {\bibfnamefont {Z.~H.}\ \bibnamefont {Loh}},\
  }\href@noop {} {\bibfield  {journal} {\bibinfo  {journal} {Acs Nano}\
  }\textbf {\bibinfo {volume} {8}},\ \bibinfo {pages} {10931} (\bibinfo {year}
  {2014})}\BibitemShut {NoStop}%
\bibitem [{\citenamefont {Wang}\ \emph {et~al.}(2015)\citenamefont {Wang},
  \citenamefont {Luo}, \citenamefont {Yabushita}, \citenamefont {Wu},
  \citenamefont {Kobayashi}, \citenamefont {Chen},\ and\ \citenamefont
  {Li}}]{Wang-2015-Ultrafast_Multi-Level_Logic_Gates}%
  \BibitemOpen
  \bibfield  {author} {\bibinfo {author} {\bibfnamefont {Y.-T.}\ \bibnamefont
  {Wang}}, \bibinfo {author} {\bibfnamefont {C.-W.}\ \bibnamefont {Luo}},
  \bibinfo {author} {\bibfnamefont {A.}~\bibnamefont {Yabushita}}, \bibinfo
  {author} {\bibfnamefont {K.-H.}\ \bibnamefont {Wu}}, \bibinfo {author}
  {\bibfnamefont {T.}~\bibnamefont {Kobayashi}}, \bibinfo {author}
  {\bibfnamefont {C.-H.}\ \bibnamefont {Chen}}, \ and\ \bibinfo {author}
  {\bibfnamefont {L.-J.}\ \bibnamefont {Li}},\ }\href@noop {} {\bibfield
  {journal} {\bibinfo  {journal} {Sci. Rep.}\ }\textbf {\bibinfo {volume}
  {5}},\ \bibinfo {pages} {8289} (\bibinfo {year} {2015})}\BibitemShut
  {NoStop}%
\bibitem [{\citenamefont {Apalkov}\ and\ \citenamefont
  {Stockman}(2012)}]{Apalkov_Stockman_PRB_2012_Strong_Field_Reflection}%
  \BibitemOpen
  \bibfield  {author} {\bibinfo {author} {\bibfnamefont {V.}~\bibnamefont
  {Apalkov}}\ and\ \bibinfo {author} {\bibfnamefont {M.~I.}\ \bibnamefont
  {Stockman}},\ }\href@noop {} {\bibfield  {journal} {\bibinfo  {journal}
  {Phys. Rev. B}\ }\textbf {\bibinfo {volume} {86}},\ \bibinfo {pages} {165118}
  (\bibinfo {year} {2012})}\BibitemShut {NoStop}%
\bibitem [{\citenamefont {Apalkov}\ and\ \citenamefont
  {Stockman}(2013)}]{Apalkov_Stockman_PRB_2013_Metal_Nanofilm_in_Ultratsrong_Fields}%
  \BibitemOpen
  \bibfield  {author} {\bibinfo {author} {\bibfnamefont {V.}~\bibnamefont
  {Apalkov}}\ and\ \bibinfo {author} {\bibfnamefont {M.~I.}\ \bibnamefont
  {Stockman}},\ }\href@noop {} {\bibfield  {journal} {\bibinfo  {journal}
  {Phys. Rev. B}\ }\textbf {\bibinfo {volume} {88}},\ \bibinfo {pages} {245438}
  (\bibinfo {year} {2013})}\BibitemShut {NoStop}%
\bibitem [{\citenamefont {Higuchi}\ \emph {et~al.}(2014)\citenamefont
  {Higuchi}, \citenamefont {Stockman},\ and\ \citenamefont
  {Hommelhoff}}]{Stockman_et_al_PhysRevLett.113_2014_HHG}%
  \BibitemOpen
  \bibfield  {author} {\bibinfo {author} {\bibfnamefont {T.}~\bibnamefont
  {Higuchi}}, \bibinfo {author} {\bibfnamefont {M.~I.}\ \bibnamefont
  {Stockman}}, \ and\ \bibinfo {author} {\bibfnamefont {P.}~\bibnamefont
  {Hommelhoff}},\ }\href@noop {} {\bibfield  {journal} {\bibinfo  {journal}
  {Phys. Rev. Lett.}\ }\textbf {\bibinfo {volume} {113}},\ \bibinfo {pages}
  {213901} (\bibinfo {year} {2014})}\BibitemShut {NoStop}%
\bibitem [{\citenamefont {Kelardeh}\ \emph {et~al.}(2014)\citenamefont
  {Kelardeh}, \citenamefont {Apalkov},\ and\ \citenamefont
  {Stockman}}]{Stockman_et_al_PhysRevB.90_2014_WS_States_of_Graphene}%
  \BibitemOpen
  \bibfield  {author} {\bibinfo {author} {\bibfnamefont {H.~K.}\ \bibnamefont
  {Kelardeh}}, \bibinfo {author} {\bibfnamefont {V.}~\bibnamefont {Apalkov}}, \
  and\ \bibinfo {author} {\bibfnamefont {M.~I.}\ \bibnamefont {Stockman}},\
  }\href@noop {} {\bibfield  {journal} {\bibinfo  {journal} {Phys. Rev. B}\
  }\textbf {\bibinfo {volume} {90}},\ \bibinfo {pages} {085313} (\bibinfo
  {year} {2014})}\BibitemShut {NoStop}%
\bibitem [{\citenamefont {Kelardeh}\ \emph
  {et~al.}(2015{\natexlab{a}})\citenamefont {Kelardeh}, \citenamefont
  {Apalkov},\ and\ \citenamefont
  {Stockman}}]{Stockman_et_al_PRB_2016_Graphene_in_Ultrafast_Field}%
  \BibitemOpen
  \bibfield  {author} {\bibinfo {author} {\bibfnamefont {H.~K.}\ \bibnamefont
  {Kelardeh}}, \bibinfo {author} {\bibfnamefont {V.}~\bibnamefont {Apalkov}}, \
  and\ \bibinfo {author} {\bibfnamefont {M.~I.}\ \bibnamefont {Stockman}},\
  }\href@noop {} {\bibfield  {journal} {\bibinfo  {journal} {Phys. Rev. B}\
  }\textbf {\bibinfo {volume} {91}},\ \bibinfo {pages} {0454391} (\bibinfo
  {year} {2015}{\natexlab{a}})}\BibitemShut {NoStop}%
\bibitem [{\citenamefont {Kelardeh}\ \emph
  {et~al.}(2015{\natexlab{b}})\citenamefont {Kelardeh}, \citenamefont
  {Apalkov},\ and\ \citenamefont
  {Stockman}}]{Stockman_PhysRevB.92_2015_Ultrafast_Control_Symmetry}%
  \BibitemOpen
  \bibfield  {author} {\bibinfo {author} {\bibfnamefont {H.~K.}\ \bibnamefont
  {Kelardeh}}, \bibinfo {author} {\bibfnamefont {V.}~\bibnamefont {Apalkov}}, \
  and\ \bibinfo {author} {\bibfnamefont {M.~I.}\ \bibnamefont {Stockman}},\
  }\href@noop {} {\bibfield  {journal} {\bibinfo  {journal} {Phys. Rev. B}\
  }\textbf {\bibinfo {volume} {92}},\ \bibinfo {pages} {045413} (\bibinfo
  {year} {2015}{\natexlab{b}})}\BibitemShut {NoStop}%
\bibitem [{\citenamefont {Kelardeh}\ \emph {et~al.}(2016)\citenamefont
  {Kelardeh}, \citenamefont {Apalkov},\ and\ \citenamefont
  {Stockman}}]{Stockman_et_al_PhysRevB.93.155434_Graphene_Circular_Interferometry}%
  \BibitemOpen
  \bibfield  {author} {\bibinfo {author} {\bibfnamefont {H.~K.}\ \bibnamefont
  {Kelardeh}}, \bibinfo {author} {\bibfnamefont {V.}~\bibnamefont {Apalkov}}, \
  and\ \bibinfo {author} {\bibfnamefont {M.~I.}\ \bibnamefont {Stockman}},\
  }\href@noop {} {\bibfield  {journal} {\bibinfo  {journal} {Phys. Rev. B}\
  }\textbf {\bibinfo {volume} {93}},\ \bibinfo {pages} {155434} (\bibinfo
  {year} {2016})}\BibitemShut {NoStop}%
\bibitem [{\citenamefont {Wismer}\ \emph {et~al.}(2016)\citenamefont {Wismer},
  \citenamefont {Kruchinin}, \citenamefont {Ciappina}, \citenamefont
  {Stockman},\ and\ \citenamefont
  {Yakovlev}}]{Stockman_Yakovlev_et_al_PhysRevLett.116_2016_Strong_Field_Dynamics_in_Semiconductors}%
  \BibitemOpen
  \bibfield  {author} {\bibinfo {author} {\bibfnamefont {M.~S.}\ \bibnamefont
  {Wismer}}, \bibinfo {author} {\bibfnamefont {S.~Y.}\ \bibnamefont
  {Kruchinin}}, \bibinfo {author} {\bibfnamefont {M.}~\bibnamefont {Ciappina}},
  \bibinfo {author} {\bibfnamefont {M.~I.}\ \bibnamefont {Stockman}}, \ and\
  \bibinfo {author} {\bibfnamefont {V.~S.}\ \bibnamefont {Yakovlev}},\
  }\href@noop {} {\bibfield  {journal} {\bibinfo  {journal} {Phys. Rev. Lett.}\
  }\textbf {\bibinfo {volume} {116}},\ \bibinfo {pages} {197401} (\bibinfo
  {year} {2016})}\BibitemShut {NoStop}%
\bibitem [{\citenamefont {Motlagh}\ \emph {et~al.}(2017)\citenamefont
  {Motlagh}, \citenamefont {Apalkov},\ and\ \citenamefont
  {Stockman}}]{Stockman_et_al_PhysRevB.95_2017_Crystalline_TI}%
  \BibitemOpen
  \bibfield  {author} {\bibinfo {author} {\bibfnamefont {S.~A.~O.}\
  \bibnamefont {Motlagh}}, \bibinfo {author} {\bibfnamefont {V.}~\bibnamefont
  {Apalkov}}, \ and\ \bibinfo {author} {\bibfnamefont {M.~I.}\ \bibnamefont
  {Stockman}},\ }\href@noop {} {\bibfield  {journal} {\bibinfo  {journal}
  {Phys. Rev. B}\ }\textbf {\bibinfo {volume} {95}},\ \bibinfo {pages} {085438}
  (\bibinfo {year} {2017})}\BibitemShut {NoStop}%
\bibitem [{\citenamefont {Kwon}\ \emph {et~al.}(2016)\citenamefont {Kwon},
  \citenamefont {Paasch-Colberg}, \citenamefont {Apalkov}, \citenamefont {Kim},
  \citenamefont {Kim}, \citenamefont {Stockman},\ and\ \citenamefont
  {Kim}}]{Stockman_et_al_Sci_Rep_2016_Semimetallization}%
  \BibitemOpen
  \bibfield  {author} {\bibinfo {author} {\bibfnamefont {O.}~\bibnamefont
  {Kwon}}, \bibinfo {author} {\bibfnamefont {T.}~\bibnamefont
  {Paasch-Colberg}}, \bibinfo {author} {\bibfnamefont {V.}~\bibnamefont
  {Apalkov}}, \bibinfo {author} {\bibfnamefont {B.-K.}\ \bibnamefont {Kim}},
  \bibinfo {author} {\bibfnamefont {J.-J.}\ \bibnamefont {Kim}}, \bibinfo
  {author} {\bibfnamefont {M.~I.}\ \bibnamefont {Stockman}}, \ and\ \bibinfo
  {author} {\bibfnamefont {D.~E.}\ \bibnamefont {Kim}},\ }\href@noop {}
  {\bibfield  {journal} {\bibinfo  {journal} {Sci. Rep.}\ }\textbf {\bibinfo
  {volume} {6}},\ \bibinfo {pages} {21272} (\bibinfo {year}
  {2016})}\BibitemShut {NoStop}%
\bibitem [{\citenamefont {Higuchi}\ \emph {et~al.}(2017)\citenamefont
  {Higuchi}, \citenamefont {Heide}, \citenamefont {Ullmann}, \citenamefont
  {Weber},\ and\ \citenamefont
  {Hommelhoff}}]{Higuchi_Hommelhoff_et_al_Nature_2017_Currents_in_Graphene}%
  \BibitemOpen
  \bibfield  {author} {\bibinfo {author} {\bibfnamefont {T.}~\bibnamefont
  {Higuchi}}, \bibinfo {author} {\bibfnamefont {C.}~\bibnamefont {Heide}},
  \bibinfo {author} {\bibfnamefont {K.}~\bibnamefont {Ullmann}}, \bibinfo
  {author} {\bibfnamefont {H.~B.}\ \bibnamefont {Weber}}, \ and\ \bibinfo
  {author} {\bibfnamefont {P.}~\bibnamefont {Hommelhoff}},\ }\href@noop {}
  {\bibfield  {journal} {\bibinfo  {journal} {Nature}\ }\textbf {\bibinfo
  {volume} {550}},\ \bibinfo {pages} {224} (\bibinfo {year}
  {2017})}\BibitemShut {NoStop}%
\bibitem [{\citenamefont
  {Houston}(1940)}]{Houston_PR_1940_Electron_Acceleration_in_Lattice}%
  \BibitemOpen
  \bibfield  {author} {\bibinfo {author} {\bibfnamefont {W.~V.}\ \bibnamefont
  {Houston}},\ }\href@noop {} {\bibfield  {journal} {\bibinfo  {journal} {Phys.
  Rev.}\ }\textbf {\bibinfo {volume} {57}},\ \bibinfo {pages} {184} (\bibinfo
  {year} {1940})}\BibitemShut {NoStop}%
\bibitem [{\citenamefont {Wilczek}\ and\ \citenamefont
  {Zee}(1984)}]{Wiczek_Zee_PhysRevLett.52_1984_Nonabelian_Berry_Phase}%
  \BibitemOpen
  \bibfield  {author} {\bibinfo {author} {\bibfnamefont {F.}~\bibnamefont
  {Wilczek}}\ and\ \bibinfo {author} {\bibfnamefont {A.}~\bibnamefont {Zee}},\
  }\href@noop {} {\bibfield  {journal} {\bibinfo  {journal} {Phys. Rev. Lett.}\
  }\textbf {\bibinfo {volume} {52}},\ \bibinfo {pages} {2111} (\bibinfo {year}
  {1984})}\BibitemShut {NoStop}%
\bibitem [{\citenamefont {Xiao}\ \emph {et~al.}(2010)\citenamefont {Xiao},
  \citenamefont {Chang},\ and\ \citenamefont
  {Niu}}]{Xiao_Niu_RevModPhys.82_2010_Berry_Phase_in_Electronic_Properties}%
  \BibitemOpen
  \bibfield  {author} {\bibinfo {author} {\bibfnamefont {D.}~\bibnamefont
  {Xiao}}, \bibinfo {author} {\bibfnamefont {M.-C.}\ \bibnamefont {Chang}}, \
  and\ \bibinfo {author} {\bibfnamefont {Q.}~\bibnamefont {Niu}},\ }\href@noop
  {} {\bibfield  {journal} {\bibinfo  {journal} {Rev. Mod. Phys.}\ }\textbf
  {\bibinfo {volume} {82}},\ \bibinfo {pages} {1959} (\bibinfo {year}
  {2010})}\BibitemShut {NoStop}%
\bibitem [{\citenamefont {Yang}\ and\ \citenamefont
  {Liu}(2014)}]{Yang_Liu_PhysRevB.90_2014_Non-Abelian_Berry_Curvature_and_Nonlinear_Optics}%
  \BibitemOpen
  \bibfield  {author} {\bibinfo {author} {\bibfnamefont {F.}~\bibnamefont
  {Yang}}\ and\ \bibinfo {author} {\bibfnamefont {R.~B.}\ \bibnamefont {Liu}},\
  }\href@noop {} {\bibfield  {journal} {\bibinfo  {journal} {Phys. Rev. B}\
  }\textbf {\bibinfo {volume} {90}},\ \bibinfo {pages} {245205} (\bibinfo
  {year} {2014})}\BibitemShut {NoStop}%
\bibitem [{\citenamefont
  {Bloch}(1929)}]{Bloch_Z_Phys_1929_Functions_Oscillations_in_Crystals}%
  \BibitemOpen
  \bibfield  {author} {\bibinfo {author} {\bibfnamefont {F.}~\bibnamefont
  {Bloch}},\ }\href@noop {} {\bibfield  {journal} {\bibinfo  {journal} {Z.
  Phys. A}\ }\textbf {\bibinfo {volume} {52}},\ \bibinfo {pages} {555}
  (\bibinfo {year} {1929})}\BibitemShut {NoStop}%
\bibitem [{\citenamefont {Yao}\ \emph {et~al.}(2008)\citenamefont {Yao},
  \citenamefont {Xiao},\ and\ \citenamefont
  {Niu}}]{Niu_et_al_PhysRevB.77_2008_Valley_Selection_in_Graphene}%
  \BibitemOpen
  \bibfield  {author} {\bibinfo {author} {\bibfnamefont {W.}~\bibnamefont
  {Yao}}, \bibinfo {author} {\bibfnamefont {D.}~\bibnamefont {Xiao}}, \ and\
  \bibinfo {author} {\bibfnamefont {Q.}~\bibnamefont {Niu}},\ }\href@noop {}
  {\bibfield  {journal} {\bibinfo  {journal} {Phys. Rev. B}\ }\textbf {\bibinfo
  {volume} {77}},\ \bibinfo {pages} {235406} (\bibinfo {year}
  {2008})}\BibitemShut {NoStop}%
\bibitem [{\citenamefont {Liu}\ \emph {et~al.}(2011)\citenamefont {Liu},
  \citenamefont {Bian}, \citenamefont {Miller},\ and\ \citenamefont
  {Chiang}}]{Chiang_et_al_PhysRevLett.107_Berry_Phase_in_Graphene_ARPES}%
  \BibitemOpen
  \bibfield  {author} {\bibinfo {author} {\bibfnamefont {Y.}~\bibnamefont
  {Liu}}, \bibinfo {author} {\bibfnamefont {G.}~\bibnamefont {Bian}}, \bibinfo
  {author} {\bibfnamefont {T.}~\bibnamefont {Miller}}, \ and\ \bibinfo {author}
  {\bibfnamefont {T.~C.}\ \bibnamefont {Chiang}},\ }\href@noop {} {\bibfield
  {journal} {\bibinfo  {journal} {Phys. Rev. Lett.}\ }\textbf {\bibinfo
  {volume} {107}},\ \bibinfo {pages} {166803} (\bibinfo {year}
  {2011})}\BibitemShut {NoStop}%
\bibitem [{\citenamefont {Xu}\ \emph {et~al.}(2013)\citenamefont {Xu},
  \citenamefont {Liang}, \citenamefont {Shi},\ and\ \citenamefont
  {Chen}}]{Xu_et_al_Chem_Rev_2013_Graphene-Like_2D_Materials}%
  \BibitemOpen
  \bibfield  {author} {\bibinfo {author} {\bibfnamefont {M.~S.}\ \bibnamefont
  {Xu}}, \bibinfo {author} {\bibfnamefont {T.}~\bibnamefont {Liang}}, \bibinfo
  {author} {\bibfnamefont {M.~M.}\ \bibnamefont {Shi}}, \ and\ \bibinfo
  {author} {\bibfnamefont {H.~Z.}\ \bibnamefont {Chen}},\ }\href@noop {}
  {\bibfield  {journal} {\bibinfo  {journal} {Chem. Rev.}\ }\textbf {\bibinfo
  {volume} {113}},\ \bibinfo {pages} {3766} (\bibinfo {year}
  {2013})}\BibitemShut {NoStop}%
\bibitem [{\citenamefont {Kelardeh}\ \emph {et~al.}(2017)\citenamefont
  {Kelardeh}, \citenamefont {Apalkov},\ and\ \citenamefont
  {Stockman}}]{Stockman_et_al_PhysRevB.96_2017_Berry_Phase}%
  \BibitemOpen
  \bibfield  {author} {\bibinfo {author} {\bibfnamefont {H.~K.}\ \bibnamefont
  {Kelardeh}}, \bibinfo {author} {\bibfnamefont {V.}~\bibnamefont {Apalkov}}, \
  and\ \bibinfo {author} {\bibfnamefont {M.~I.}\ \bibnamefont {Stockman}},\
  }\href@noop {} {\bibfield  {journal} {\bibinfo  {journal} {Phys. Rev. B}\
  }\textbf {\bibinfo {volume} {96}},\ \bibinfo {pages} {075409} (\bibinfo
  {year} {2017})}\BibitemShut {NoStop}%
\bibitem [{\citenamefont
  {Winkle.}(2003)}]{Winkler_Springer_Berlin_Heidelberg_2003_Acceleration_of_Electrons_in_a_Crystal_Lattice}%
  \BibitemOpen
  \bibfield  {author} {\bibinfo {author} {\bibfnamefont {R.}~\bibnamefont
  {Winkle.}},\ }\href {\doibase 10.1007/b13586} {\bibfield  {journal} {\bibinfo
   {journal} {Springer Berlin Heidelberg}\ ,\ \bibinfo {pages} {61}} (\bibinfo
  {year} {2003})}\BibitemShut {NoStop}%
\bibitem [{\citenamefont
  {Butcher}(1963)}]{butcher_Cambridge_University_Press_1963_Runge_Kutta_integration_processes}%
  \BibitemOpen
  \bibfield  {author} {\bibinfo {author} {\bibfnamefont {J.~C.}\ \bibnamefont
  {Butcher}},\ }\href {\doibase 10.1017/S1446788700027932} {\bibfield
  {journal} {\bibinfo  {journal} {Journal of the Australian Mathematical
  Society}\ }\textbf {\bibinfo {volume} {3}},\ \bibinfo {pages} {185–201}
  (\bibinfo {year} {1963})}\BibitemShut {NoStop}%
\end{thebibliography}

%

\end{document}